\documentclass[twocolumn]{aastex631}

\newcommand{\Xlan}{X$_{\rm lan}$}

\newif\ifcomments
\commentstrue 

\usepackage{amsfonts,amsmath,natbib,subfigure,color,gensymb,longtable,verbatim,graphicx,float,soul,multirow}

\begin{document}

\title{Non-Thermal Ionization of Kilonova Ejecta: Observable Impacts}

\author[0000-0001-6415-0903]{D. Brethauer}
\affiliation{Department of Astronomy, University of California, Berkeley, CA 94720-3411, USA}
\affiliation{Berkeley Center for Multi-messenger Research on Astrophysical Transients and Outreach (Multi-RAPTOR), University of California, Berkeley, CA 94720-3411, USA}

\author[0000-0002-5981-1022]{D. Kasen}
\affiliation{Department of Physics, University of California, 366 Physics North MC 7300, Berkeley, CA 94720, USA}
\affiliation{Department of Astronomy, University of California, Berkeley, CA 94720-3411, USA}
\affiliation{Nuclear Science Division, Lawrence Berkeley National Laboratory, 1 Cyclotron Rd, Berkeley, CA, 94720, USA}

\author[0000-0003-4768-7586]{R. Margutti}
\affiliation{Department of Astronomy, University of California, Berkeley, CA 94720-3411, USA}
\affiliation{Department of Physics, University of California, 366 Physics North MC 7300, Berkeley, CA 94720, USA}
\affiliation{Berkeley Center for Multi-messenger Research on Astrophysical Transients and Outreach (Multi-RAPTOR), University of California, Berkeley, CA 94720-3411, USA}

\author[0000-0002-7706-5668]{R. Chornock}
\affiliation{Department of Astronomy, University of California, Berkeley, CA 94720-3411, USA}
\affiliation{Berkeley Center for Multi-messenger Research on Astrophysical Transients and Outreach (Multi-RAPTOR), University of California, Berkeley, CA 94720-3411, USA}



\begin{abstract}

The characteristic rapid rise and decline at optical wavelengths of a kilonova is the product of the low ejecta mass ($\lesssim 0.05 M_\odot$) and high ejecta velocity ($\gtrsim 0.1$c). We show that, even at very early times ($\lesssim 2$ days), regions of ejecta fall below critical density and temperature thresholds at which non-local thermodynamic equilibrium (NLTE) effects become important. Here, we present an approximate method for calculating the ionization state of the ejecta that accounts for the NLTE impact of high-energy electrons produced in the beta decay of freshly synthesized $r$-process elements. We find that incorporating ionization from high-energy electrons produces an ``inverted" and ``blended" ionization structure, where the most highly ionized species are located in the fastest moving homologous ejecta and multiple ionization states coexist. In radiation transport calculations, the higher degree of ionization reduces line blanketing in optical bands, leading to improved agreement with the light curve properties of AT\,2017gfo such as the duration, decay rates, brightness, and colors. Our quasi-NLTE implementation helps to alleviate tensions in kilonova modeling: for high-velocity ($\sim 0.3c$) ejecta components our models require less mass for a given peak brightness in optical bands, by as much as a factor of 3; our models can explain the presence of observed features associated to Sr II, W III, Se III, and Te III under conditions where LTE models would predict only neutral species; and we naturally predict the coexistence of species like Sr II and Ce III without the need for fine-tuning of the ejecta properties.

\end{abstract}



\section{Introduction} \label{sec:intro}

Compact-object mergers involving a neutron star (NS) have long been at the forefront of high-energy phenomena, from short gamma ray bursts (GRBs; e.g., \citealt{Eichler89,Narayan92}) to sites of \textit{rapid neutron capture} ($r$-process) nucleosynthesis \citep{Lattimer&Schramm74,Lattimer&Schramm76,Symbalisty&Schramm82,Eichler89,Freiburghaus99,Rosswog99}. The discovery of the gravitational wave source GW\,170817 and its electromagnetic counterparts, AT\,2017gfo and GRB\,170817A \citep{Abbott17,Andreoni17,Arcavi17,Chornock17,Coulter17,Cowperthwaite17,Diaz17,Drout17,Evans17,Goldstein17,Hu17,Kasliwal17,Lipunov17,Pian17,Savchenko17,Shappee17,Smartt17,Soares-Santos17,Tanvir17,Utsumi17,Valenti17,Pozanenko18}, confirmed the connection between compact-object mergers and some short GRBs, ushering in a new era transient multi-messenger astronomy with gravitational waves (see \citealt{Margutti&Chornock21} and \citealt{Nakar20} for reviews). The kilonova associated with GW\,170817, AT\,2017gfo, was a multi-wavelength transient powered by the radioactive decay of $r$-process material. The detected emission of AT\,2017gfo lasted between $\sim$ days and $\sim$ weeks depending on the observed wavelength (e.g., data compilation by \citealt{Villar17} and references therein). The modeling of the light curves and spectra of AT\,2017gfo has been used to extract information about the heavy-metal content (typically denoted as the mass fraction of lanthanides, \Xlan, a product of $r$-process nucleosynthesis), as well as the mass and velocity of the different ejecta components. Accurate estimates of the mass, velocity, and \Xlan\, are crucial because the properties of kilonova ejecta can be used to understand the fate of the merger remnant (e.g., \citealt{Margalit17,Radice20,Radice23}), which constrains the neutron star equation of state, and constrains the contribution of kilonovae to the evolution of the heavy-metal content of the Universe (e.g., \citealt{Hotokezaka15,Qian&Wasserburg07,Wallner15,Ji16,Kasen17,Rosswog18}). 

The light curves and spectra are most strongly determined by the wavelength-dependent opacity of the lanthanide elements and their abundance in the ejecta. The valence $f$-shell electrons of lanthanide elements lead to optical and  ultraviolet (UV) opacities $\gtrsim 10^2$ cm$^2$ g$^{-1}$, which strongly reprocess optical and UV light into infrared emission (IR; e.g., \citealt{Kasen13,Tanaka20,Fontes20}). Despite the importance to kilonova observables, the high complexity of lanthanide atomic structure and lack of direct experimental constraints remain a hurdle to kilonova modeling. As a result, many kilonova radiative simulations make some combination of simplifying assumptions from two broad categories: i) physics-based assumptions or ii) phase-space assumptions. Physics-based assumptions range from using a wavelength-independent (gray) or analytical prescriptions for the opacity (e.g., \citealt{Metzger17, Bulla19}) to assuming Local Thermodynamic Equilibrium (LTE; e.g., \citealt{Wollaeger18,Tanaka20,Bulla23,Shingles23,Brethauer24}). Models that implement phase-space assumptions incorporate NLTE effects, but due to the computational cost of full-scale NLTE restrict the number of parameters involved in the calculation such as reducing the dimensionality of the ejecta (e.g., \citealt{Pognan23}), or narrowing the number of chemical elements in the ejecta (e.g., \citealt{Hotokezaka21,Pognan22b,Banerjee25}). 

While there were many predictions about kilonovae that were confirmed with AT\,2017gfo and to a lesser extent by other kilonova observations (e.g., \citealt{Berger13,Tanvir13,Rastinejad22,Levan24}) such as spectral peak, light curve duration, and IR brightness, there are still many broad discrepancies between model predictions and observations. Most notably, the color evolution and optical light curve decay rates of LTE kilonova models drastically deviate from the light curve of AT\,2017gfo (e.g., Figs. 9--10 of \citealt{Brethauer24}). These models typically overpredict IR fluxes and fade too rapidly at optical wavelengths. Yet, the models produce \emph{bolometric} luminosity outputs that are consistent with observations. Additionally, these discrepancies occur across a wide variety of models using different atomic datasets and radiative transfer codes, while being absent in gray-opacity models (see various models within \citealt{Villar17,Wollaeger18,Tanaka20,Bulla23,Brethauer24}). Taken together, these modeling discrepancies may point to a systematic overestimation of wavelength-dependent opacities in LTE models.

One effect that can reduce the optical opacity is an increase in the degree of ionization. The tight energy-level spacing of neutral species leads to strong line blanketing in the optical, whereas in high ionization states the line blanketing shifts towards shorter wavelengths, into the UV. Non-thermal sources of ionization, such as ionization from high-energy electrons produced via beta-decay of $r$-process elements, have been shown to play a significant role in determining the ionization structure of kilonova ejecta (e.g., \citealt{Hotokezaka21,Pognan22b}) and SN ejecta (e.g., \citealt{Jerkstrand17}), driving departures from LTE. Crucially, if these effects become significant near and above the photosphere where the spectrum is forming, the resulting spectra may be altered. For example, when the outer ejecta cool to $T \lesssim 2500$ K, LTE predicts mostly neutral species, but non-thermal ionization may keep the gas partially ionized. Indeed, \cite{Kawaguchi21} have shown that as a proxy for NLTE effects, artificially removing the opacity contributions of neutral species can lead to brighter optical light curves.

While solving the full NLTE rate equations for all energy levels of lanthanides would be computationally infeasible for the timescales necessary, we present a simplified framework to compute kilonova spectra into nebular phases including time-dependent NLTE ionization effects driven by radioactivity. We find that accounting for NLTE effects has significant effects on the broadband light curves and can help to resolve the discrepancies between observations and models.  

The paper is organized as follows: in \S \ref{Sec:QNLTE} we discuss the importance of including radioactive ionization. In \S \ref{sec:methods} we discuss the radiative transfer code \texttt{Sedona} we employ for our models, the set up of our models, and underlying equations. We then explore the ionization structure and light curves that result from NLTE effects in \S \ref{Sec:IonStruct}. We then discuss the impacts of these results on modeling and their limits in \S \ref{Sec:Discussion}. Finally, we conclude the paper in \S \ref{Sec:Conc}.

\section{Quasi-NLTE Model} \label{Sec:QNLTE}

\subsection{Non-Thermal Ionization}
\label{subsec:formulae}
In order for NLTE effects to become significant, the NLTE ionization rates must be at least comparable to the rates from thermal effects. Considering two adjacent ionization states $i$ and $i+1$, the NLTE rate equations determining the ratio in steady-state are 
\begin{equation}
    n_i ({R_{\rm PI} + R_{\rm nt}})
    = n_{i+1}  {R_{\rm rec}}
    \label{eq:rate_equation}
\end{equation}

\noindent where $R_{\rm PI}$ is the photoionization rate, $R_{\rm nt}$ is the  non-thermal ionization rate, $R_{\rm rec}$ is the total recombination rate, and $n_i$ is the number density of the $i$th ionization species. The photoionization rate is defined by
\begin{equation}
R_{\rm PI} = 
4\pi \int_{\nu_t}^\infty \frac{J_\nu}{h \nu} \sigma_{\rm PI}(\nu)~d \nu
\label{eq:photoionization_rate}
\end{equation}
\noindent where $J_\nu$ is the local radiation field, $\sigma_{\rm PI}$ ($\nu$) is the frequency-dependent photoionization cross section, $h$ is the Planck constant, and $\nu_t$ is the threshold frequency of a photon corresponding to the ionization energy of the ion. We neglect the effects of thermal collisional ionization as the rates are much weaker than photoionization for kilonova-like conditions ($\rho \lesssim 10^{-10} {\rm g\,cm^{-3}}$, $T \sim$ few $\times 10^3$ K). 

The non-thermal ionization rate we focus on originates from the high-energy electrons released from the beta-decay of freshly synthesized $r$-process material (``radioactive ionization"), described by 

\begin{equation} \label{eq:rad}
R_{\rm nt} = f_{\rm ion} \frac{\langle A \rangle m_p f_{\rm \beta} \dot{\epsilon}}{\chi}
\end{equation}

\noindent where $\dot{\epsilon}$ is the $r$-process heating rate per unit mass, $f_{\beta}$ is the fraction of the $r$-process heating rate carried by high-energy electrons which thermalize, $f_{\rm ion}$ is the fraction of energy from high-energy electrons that goes into ionization (see \S\ref{subsubsec:fion}), $\langle A \rangle m_p$ is the mean atomic mass of the surrounding material, and $\chi$ is the ionization energy of the ion.

\begin{figure}[t]
    \centering
    \includegraphics[width=0.47\textwidth,trim={0cm 0cm 0cm 0cm},clip]{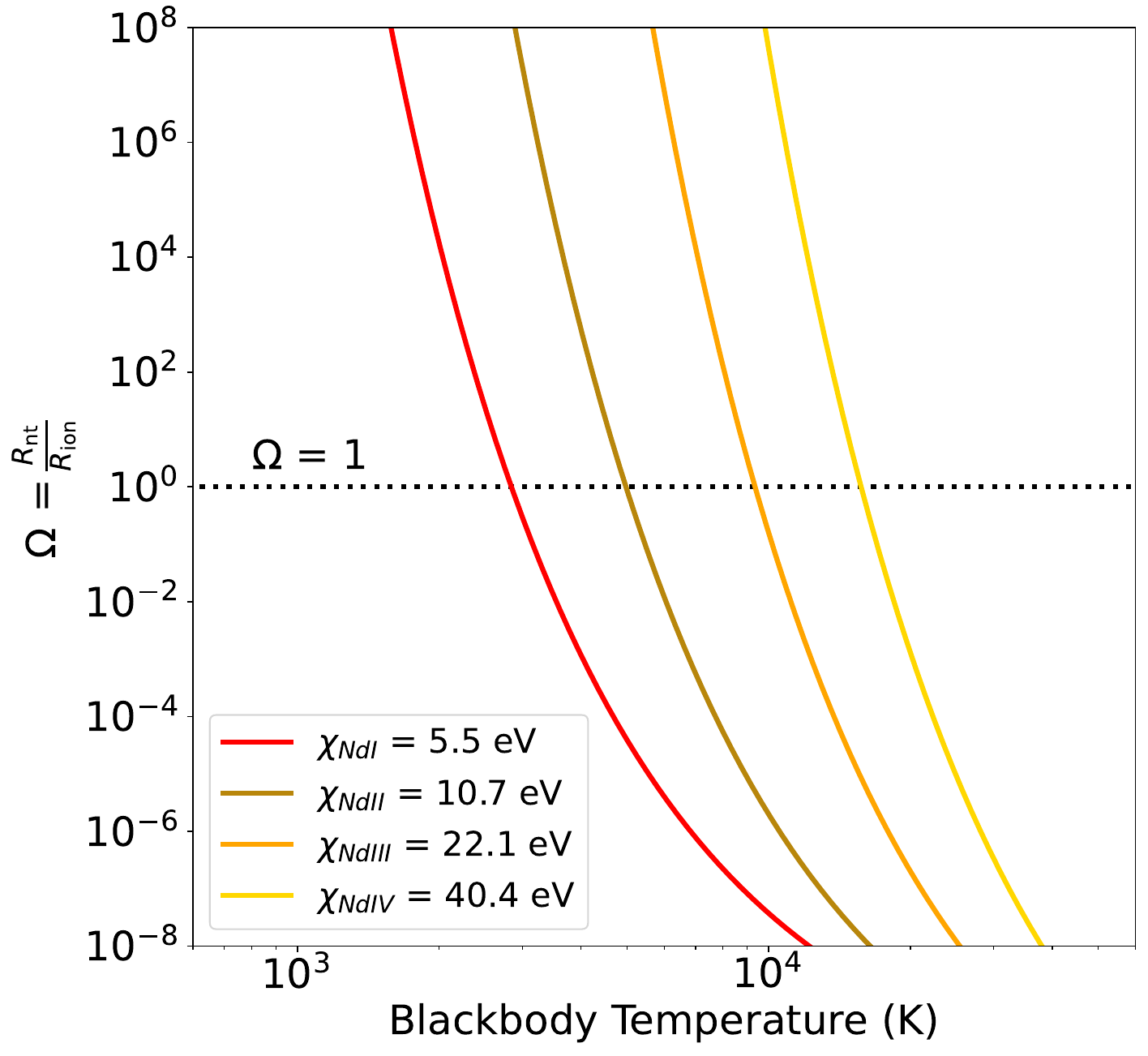}
  \centering

    \caption{The ratio of non-thermal ionization to photoionization rates ($\Omega$) for an assumed blackbody radiation spectrum of different temperatures for a fiducial zone at t = 5 days and $\rho$ = $8\times10^{-15}$ g cm$^{-3}$ for neodymium ionization energies (red to yellow lines). When $\Omega > 1$, non-thermal radioactive ionization from high-energy beta particles becomes dominant over photoionization. Radioactive ionization becomes important for neutral states at approximately 3000\,K, which is roughly the color temperature that AT\,2017gfo stalled at beginning around 4 days post explosion \citep{Drout17}. Below this temperature, radioactive ionization determines the ionization state. 
    }
    \label{Fig:OmegaIonComp}

\end{figure}

A useful metric to quantify the importance of non-thermal ionization is the ratio of non-thermal ionization to photoionization
\begin{equation}
\Omega = \frac{R_{\rm nt}}{R_{\rm PI}} 
\end{equation}
For $\Omega \ll 1$, the radiation field dominates and non-thermal ionization can be ignored. If the radiation field is approximately blackbody (as is often the case in kilonovae) the ionization state will tend towards the LTE value predicted by the Saha equation.  For $\Omega \gg 1$, the ionization is set by non-thermal effects and may deviate drastically from the LTE prediction.  

Figure \ref{Fig:OmegaIonComp} shows the value of $\Omega $ taking the radiation field to be a blackbody for kilonova-like conditions. The photoionization rate $R_{\rm PI}$ drops exponentially when the blackbody temperature drops below $kT_r \approx \chi$, where $k$ is the Boltzmann constant, because there are exponentially fewer ionizing photons above the threshold frequency. 

We define a critical temperature $T_{\rm nt}$ below which non-thermal effects should dominate the ionization. To obtain a rough analytical estimate of the critical temperature, we approximate the photoionization rate and solve for when $\Omega = 1$. The photoionization rate is dominated by photons near the threshold frequency, $\nu_t$, due to the photoionization cross-section  declining $\sim\nu^{-3}$, so we evaluate the integrand in Equation~\ref{eq:photoionization_rate} at $\nu_t$. Assuming the radiation field is a blackbody and $h\nu_t \gg kT$, we find
\begin{equation}
T_{\rm nt} = \frac{ \chi/k}{\ln \xi }
\approx 2300
\left( \frac{\chi}{5~{\rm eV}} \right)
\left( \frac{25}{\ln \xi} \right)~{\rm K}
\end{equation}
\noindent where
\begin{equation}
\xi \equiv  \left[  \frac{\chi} {  f_{\rm ion} \langle A \rangle m_p f_{\rm \beta} \dot{\epsilon}} 
\frac{ 8 \pi\sigma_t   \nu_t^3}{c^2} \right]
\end{equation}
\noindent where $\sigma_t$ is the value of the photoionization cross section at the threshold frequency. As the dependence on $\xi$ is only logarithmic, $T_{\rm nt}$ is set mainly by the ionization energy, $\chi$. 

At early times, when the ejecta are hot and above $T_{\rm nt}$, photoionization dominates and the ionization state will be approximately that of LTE if the radiation field is near blackbody. Since $T_{\rm nt}$ is roughly the same as the temperature at which the Saha equation predicts that the ion should recombine, non-thermal effects will become substantial when LTE predicts the matter will decrease in ionization.  

\subsection{Recombination and Non-Thermal Ionization State} \label{subsec:ionstate}

We now consider sources of recombination and the resulting ionization states from the combined ionization and recombination rates. When $T < T_{\rm nt}$ (and therefore $R_{\rm nt} > R_{\rm PI}$), non-thermal effects determine the ionization state and from Eq.~\ref{eq:rate_equation} the ionization ratio will be 
\begin{equation}
\frac{n_{i+1}}{n_i} = \eta_i~~~~{\rm where}
~\eta_i  \equiv  \frac{R_{\rm nt}}{R_{\rm rec} }
\label{eq:eta_definition}
\end{equation}

The recombination rate can be approximated as
\begin{multline} \label{eq:recomb}
R_{\rm rec} = f_e n\alpha_i = 3\times10^{-5}\, {\rm s^{-1}} \left( \frac{i}{1} \right)^2 \left( \frac{f_e}{1}\right)\times \\ \left( \frac{\alpha_0}{3\times10^{-13} {\rm cm^{3} s^{-1}}} \right)\left(\frac{n}{10^8 {\rm cm^{-3}}} \right) \left( \frac{T}{10^4{\rm K}} \right)^{-3/4} 
\end{multline}

\noindent as per the prescription from \cite{Axelrod80} for radiative recombination of the $i$th ionization state, where $n$ is the total number density of the ejecta of all species, and $f_e$ is defined by the free electron density, $n_e$, and $n$ by $f_e\times n \equiv n_e$. Notably, the recombination rate we use includes ionization from the ground state, as the lowest density ejecta may have a sufficiently low optical depth to allow the photoionizing photons to escape. For lanthanide elements, we set $\alpha_0$ = $3\times10^{-12}$ cm$^3$ s$^{-1}$ (for more details, see Appendix \ref{App:Recomb}).

\begin{figure}[t]
  \centering
    
    \includegraphics[width=0.48\textwidth,trim={0cm 0cm 0cm 0cm},clip]{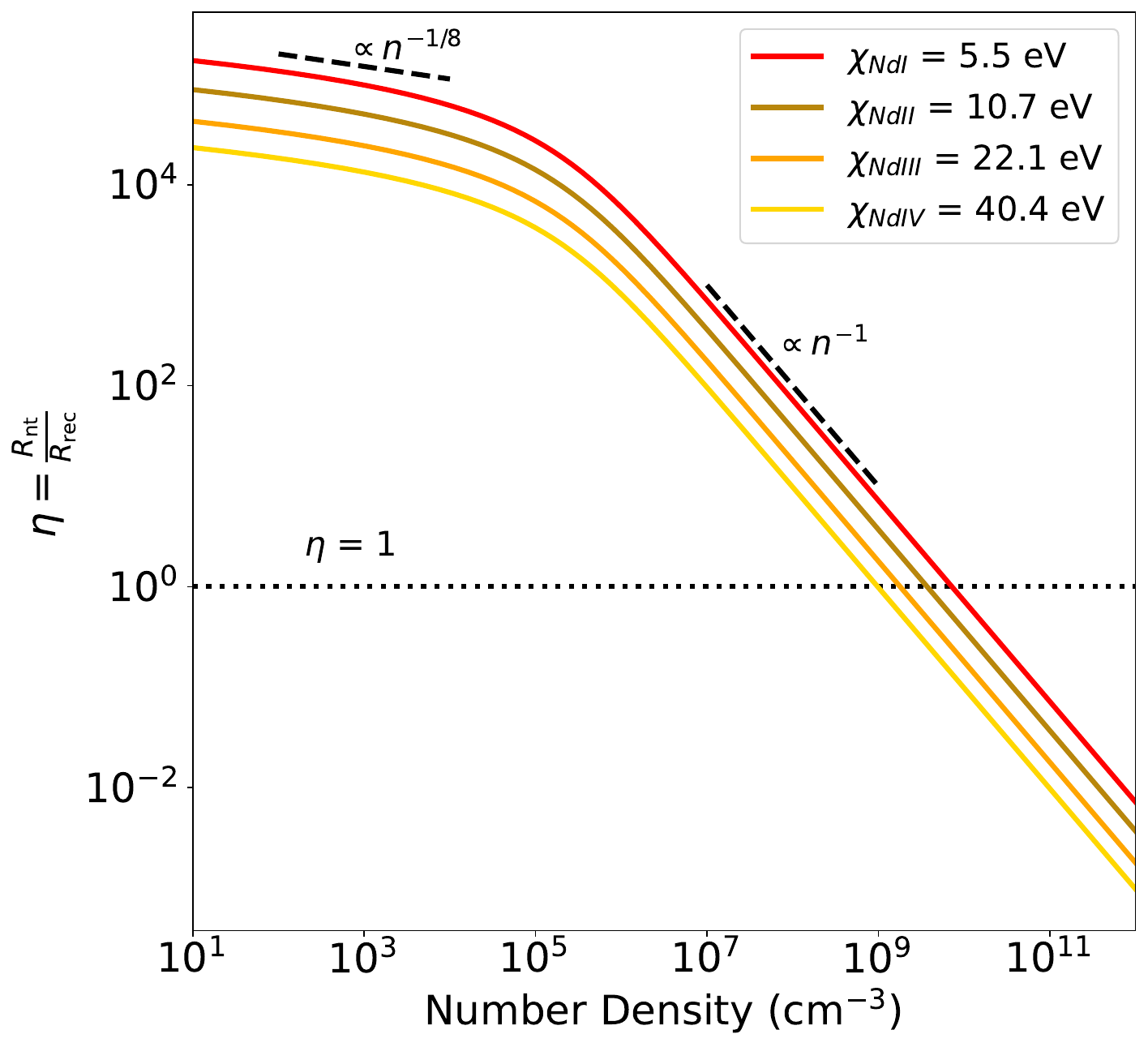}
    \caption{The ratio of non-thermal ionization to recombination, $\eta$,  assuming $T$ = 5000 K at t = 5 days, $f_{\rm ion} = \frac{1}{3}$, $f_e = 1$, $\langle A \rangle = 80$, and $f_\beta \times\dot{\epsilon}$ is the heating rate used in \texttt{Sedona} (see \S \ref{Subsec:rproc}) convolved with the local thermalization prescription (see \citealt{Brethauer24} for more details) for neodymium ionization energies. The color scheme is the same as Fig. \ref{Fig:OmegaIonComp}. At early times, the ejecta are characterized by high densities, then the ejecta expand with time, leading to a linear increase of the value of $\eta$ as the density decreases. At high number densities, $n$, the ionization ratio drops as $\eta \propto n^{-1}$, due to the increasing recombination rate. At low densities beta particles no longer thermalize efficiently and $\eta$ depends weakly on density. } 
    
    \label{Fig:EtaIonComp}
\end{figure}

Figure \ref{Fig:EtaIonComp} shows how $\eta$ evolves as a function of number density of the ejecta for kilonova-like conditions. At early times when the ejecta are dense, $\eta < 1$ and non-thermal ionization cannot efficiently ionize the ejecta due to the high recombination rate. As time progresses and the density decreases, $\eta$ increases linearly as radioactive ionization becomes more dominant until the density becomes sufficiently low that the energy from $r$-process decay no longer thermalizes completely, and $f_\beta$ decreases, causing $\eta$ to rise less steeply. For a given $r$-process heating rate and temperature, we define a critical density below which $\eta > 1$ and the ejecta will become ionized due to non-thermal ionization

\begin{equation}
n_{\rm nt} = \frac{  f_{\rm ion} \langle A \rangle m_p f_{\rm \beta} \dot{\epsilon}}{\chi f_e\alpha_{i+1}} 
\end{equation}

For typical kilonova values, $n_{\rm nt}$ can be approximated as

\begin{multline}
    n_{\rm nt} \approx 3.3\times10^{9} \left( \frac{\chi}{5 {\rm\, eV}} \right)^{-1} \left( \frac{f_{\rm ion}}{0.1} \right) \left( \frac{f_{\rm \beta}}{0.1} \right) \left( \frac{f_{\rm e}}{1} \right)^{-1} \\\times\left( \frac{i}{1} \right)^{-2} \left( \frac{\dot{\epsilon}}{10^{10} {\rm erg\, g^{-1} s^{-1}}} \right) \left( \frac{\langle A \rangle}{80} \right) \left( \frac{T}{5000 {\rm K}} \right)^{3/4} {\rm cm^{-3}}
\end{multline}

Now that we have a quantitative understanding of how $\eta$ is expected to behave, we examine how $\eta$ will evolve in time. As the radioactive ionization is driven by thermalization of high-energy electrons, the time dependence of $\eta$ will arise from the $r$-process heating rate convolved with the thermalization efficiency of those high-energy electrons. As \cite{Kasen19} showed, the thermalization efficiency can approximated as 

\begin{equation}
f_{\rm \beta,th} \approx (1 + t/t_{\rm \beta, th})^{-n}
\end{equation}
\noindent where $n \approx 1-1.5 $ and $t_{\rm \beta,th}$, the electron thermalization timescale, depends on the intrinsic properties of the ejecta like mass and velocity. Thus, $f_{\rm \beta,th}\sim 1$ at early times, then $f_{\rm \beta,th} \approx (t/t_{\rm \beta, th})^{-n}$ at late times.

The $r$-process heating rate is commonly described by a power law, $\dot{\epsilon}(t) \propto t^{-1.3}$ (e.g., \citealt{Metzger10}). Combined, the radioactive ionization rate at late times will evolve as $R_{\rm rad} (t) \propto t^{-n-1.3}$. The density, recombination coefficient, and temperature evolve in time, with $\rho(t) \propto t^{-3}$ from homologous expansion and  $T(t) \propto t^{-\gamma}$ where $\gamma \sim 0.5-0.8$ from previous \texttt{Sedona} simulations and modeling of AT\,2017gfo \citep{Drout17,Waxman18,Sneppen23b}. Thus, for typical kilonova values, $\eta(t)\propto t^{(1.7-n-0.75\gamma)} \sim t^{0 \pm 0.2}$, indicating that $\eta$ will settle to a relatively constant value and the ionization state will remain roughly constant (which we explore further and show numerical results for in \S \ref{Sec:IonStruct}). \cite{Hotokezaka21} reached a similar conclusion, finding that their normalized heating rate evolved as $t^{0.2}$ at late times. 

At early times when $f_{\rm \beta,th} \sim 1$, $\eta$ is no longer dependent on $n$ and $\eta(t) \propto t^{1.7-0.75\gamma} \sim t^{1.2\pm0.1}$. Thus, $\eta$ increases with time as a result of the decreasing density and the ionization state of the ejecta increases.

Putting the equations from this section together, we can form a cohesive qualitative story of how the ejecta  evolve. Initially, the ejecta that are above $n_{\rm nt}$ are hot and dense and will be approximately in LTE. The ejecta will continue to expand and cool, eventually dropping below $T_{\rm nt}$ and $n_{\rm nt}$ where radioactivity can effectively ionize the ejecta above the LTE expectation. The deviation from LTE will be especially significant in regions of lower density, like the outermost layers, where the recombination rate is lower. As the ejecta continue to expand, the temporal evolution of the heating rate convolved with the thermalization efficiency will scale approximately similarly as the recombination rate, ``freezing" the ionization state. If the physical conditions of the ejecta were such that $n < n_{\rm nt}$ even at early times, then the ionization state of a given atomic species will increase as the ejecta expand until the beta particles no longer thermalize efficiently, at which point the ionization state will either decrease or freeze depending on how quickly the temperature decreases. If the ionization state decreases due to rapid temperature evolution, then there will be a peak-like structure in the ionization state as a function of time where the ionization state is temporarily elevated.

\section{Methods} \label{sec:methods}

\subsection{Sedona and Initial Conditions Setup} \label{sec:Sedona}

We use the Monte Carlo Radiative transfer code \texttt{Sedona} \citep{Kasen06,Roth15} to generate synthetic spectra, from which we derive light curves. Each model is spherically symmetric and defined by the total mass, the characteristic velocity of the ejecta, density profile, and the mass fraction of lanthanides. We explore the effects of each ejecta parameter by altering one characteristic at a time from the fiducial model of $M$ = 0.03 M$_\odot$, $v_k = 0.1c$, and \Xlan\, = $10^{-2}$ where $M$ is the total ejecta mass, $v_k$ is the characteristic velocity of the ejecta defined as $v_k = \sqrt{\frac{2E_k}{M}}$ and $E_k$ is the total kinetic energy of the ejecta, and $c$ is the speed of light. For the remainder of this work, we use fiducial values for our models unless otherwise specified.

We run each simulation in 1D with 80 zones that are expanding homologously with phyiscal conditions defined by a temperature, density, velocity, and chemical composition. We initialize the model at time $t_0$= 0.25\,d after the merger, early enough that radiative diffusion will not yet have caused substantial energy loss.
\texttt{Sedona} homologously expands the system thereafter. 
Homologous expansion typically sets in on  timescales well before we initialize each simulation ($\approx$ 1 s to $\sim$ hours; e.g., \citealt{Grossman14,Rosswog14,Sippens25}).

We use an ejecta density profile of a broken power law

\begin{equation}
        \rho (v,t) =
        \Biggl\{ \begin{array}{ll}
            C_\rho \dfrac{M}{v_t^3t^3} \left(\frac{v}{v_t}\right)^{-\delta} & v \leq v_t \\
            \\
             C_\rho \dfrac{M}{v_t^3t^3} \left(\frac{v}{v_t}\right)^{-q} & v > v_t, 
        \end{array} 
\end{equation}
\noindent where $\delta$ and $q$ are the power law index of the inner and outer ejecta, respectively, $v_t$ is the velocity at which the transition between the two power law indices occurs, $t$ is time, and $C_\rho$ is the normalization constant. Following \cite{Kasen17} we adopt the broken power-law density profile with $\delta = 1$ and $q = 10$, which is similar to accretion disk wind models (e.g., the velocity distributions studied in \citealt{Fryer24}) and supernova ejecta from a massive compact star such as a Wolf-Rayet (e.g., \citealt{Chevalier89}).
The transition velocity, $v_{t}$, is defined by

\begin{equation}
    v_t \equiv C_v v_k = C_v \sqrt{\frac{2E_k}{M}}, 
\end{equation}
\noindent where $C_v$ is the normalization constant to ensure the ejecta have total kinetic energy $E_k$. The maximum velocity of each model is chosen to be 3$v_t$ (2 $v_t$) for models with $v_k <$ 0.3c ($v_k =$ 0.3c) The steep $q = 10$ power law decline makes the ejecta mass above this maximum velocity negligible.

The composition of each model is based on solar abundance patterns (or meteoric where solar abundances are not available) presented in \cite{Asplund09}, and $r$-process residuals from \cite{Simmerer04} for elements with atomic number Z = 31--70. The composition is then normalized by mass fraction such that all elements of Z = 58--70 have a total mass fraction of \Xlan \, and all other elements have a total mass fraction of 1-\Xlan. We do not consider any elements of Z $\geq$ 71.

Each spectrum is calculated every 0.1\,d starting from t = 0.25\,d to 35\,d with 1524 logarithmically spaced frequency points between $10^{13}$ and $2 \times 10^{16}$ Hz. We limit hydrodynamical time steps to the minimum of 5\% of the elapsed time and 0.5\,d, which is sufficient to resolve the expansion evolution of the ejecta. The simulation is initialized assuming LTE, and then switches to QNLTE after the first hydrodynamical step. We do not include the physics of free neutron decay or shock-powered emission in order to isolate the effects of radioactive ionization, and so caution about model accuracy at times t $\lesssim$ 0.5\,d.

At each time step, Monte Carlo packets (effectively bundles of photons of a given wavelength that total up to a specified energy amount) are released in accordance to the $r$-process heating rate convolved with the thermalization efficiency and interact with a zone through scattering and absorption. Monte Carlo photons that reach the outer edge of the simulation (defined by the maximum velocity) escape the ejecta and are collected and binned in time and frequency to generate the spectral time series of the model, with all relevant Doppler shift and light travel-time effects taken into account for an observer infinitely far away.

\subsection{QNLTE}

\subsubsection{Ionization Fractions}

To solve for the ionization state of the ejecta, we define the variable $\psi_i$ that combines the non-thermal ionization rate, photoionization rate, recombination coefficient (Equations \ref{eq:photoionization_rate},\ref{eq:rad},\ref{eq:recomb}), and total number density
\begin{equation}
\psi_i = \frac{R_{\rm nt,i} + R_{\rm PI,i}}{n\alpha_{i+1}}
\end{equation}

\noindent Within \texttt{Sedona}, we calculate the photoionization rate by integrating the radiation field from Monte Carlo transport and assuming hydrogen-like photoionization cross-sections. At early times while photoionization is relevant ($\Omega \ll 1$), the radiation field is well approximated by a blackbody and we set $J_\nu = B_\nu$ where $B_\nu$ is the blackbody function. To ensure the validity of this assumption, we ran simulations assuming $J_\nu = B_\nu$ and using the true radiation field, and the models were nearly identical except for a slight increase in noise for the model where the actual radiation field was used.

Calculating the value of $\psi_i$ for all ion ratios from i = 0 (i.e., neutral) to the maximum ion $i=N$ (i.e., fully ionized) allows us to find the general solution for the fraction of an element in ionization state $i$, $f_i$,

\begin{equation} \label{Eq:IonFrac}
    f_i = \frac{f_e^{N-1-i}\prod_{j<i} \psi_j}{\sum_i \left[ f_e^{N-1-i}\prod_{j<i} \psi_j \right]}
\end{equation}

\noindent \texttt{Sedona} iteratively solves for $f_e$ using charge conservation, balancing the number of free electrons with the number and charge of ions.

We set the level populations of each species  to their LTE values (hence ``quasi"-NLTE).  Collisional excitation by non-thermal beta particles can drive deviations from LTE \citep{Pognan22b}, however this effect should be modest as long as radiative transitions between bound-bound levels dominate over the non-thermal excitations. This is likely to be the case during most of the optically thick photospheric phase. Unlike the photoionization rate, which drops sharply at low temperatures due to a dearth of UV photons, the relevant radiative bound-bound transitions occur at longer wavelengths where there is a greater number of photons. Full NLTE calculations of the level populations will be needed to improve the fidelity of our models, especially in the nebular phase when the ejecta become optically thin and the radiation field deviates substantially from a blackbody.

\subsubsection{Ionization Energy Fraction}\label{subsubsec:fion}

The fraction of radioactive energy that goes into non-thermal ionization, $f_{\rm ion} $, changes with time as a function of the ejecta ionization state.
In highly ionized media with high free electron densities, the energy of beta particles is dissipated mostly through Coulomb interactions with thermal free electrons, which \ \ suppresses $f_{\rm ion}$. Conversely, at low ionization states and low electron densities, energy is dissipated mostly through ionizations and excitations of bound electrons, increasing  $f_{\rm ion}$. Thus, $f_{\rm ion}$ helps to keep the ejecta in a intermediate ionized state.

To estimate $f_{\rm ion}$ we consider the stopping power (i.e., energy lost by a charged particle per unit distance) of electrons due to ionization and excitation given by the Bethe formula (including relativistic corrections, \citealt{Hotokezaka21}):
\begin{multline} \label{Eq:bethe}
  \frac{dE}{dx}_{\rm ion,ex} \approx \frac{4\pi Ze^4}{m_e v^2} \Biggl[ \ln\left( \frac{1}{\langle I \rangle}\sqrt{\frac{m_ev^2 E_k}{2(1-\beta^2)}} \right) \\ -\left(\sqrt{1-\beta^2} -\frac{1-\beta^2}{2}\right) \ln(2)+\frac{1-\beta^2}{2} \\ + \frac{1}{16}\left(1-\sqrt{1-\beta^2} \right)^2\Biggr]
\end{multline}
\noindent where $\beta = v/c$.
This can be compared to the stopping power of thermal plasma due to Coulomb interactions with free electrons, as derived in \cite{Bohr13}:
\begin{equation}
    \frac{dE}{dx}_{\rm elec} = \frac{4\pi e^4 f_e}{m_e v^2} \ln \left( \frac{1.123 m_e v^3}{e^2 \omega_p} \right)
\end{equation}
\noindent where $Z$ is the charge of the ion being interacted with, $v$ and $E_k$ are the velocity and the kinetic energy of the high-energy electron, respectively, $\langle I \rangle$ is the average ionization potential (where we use values from \texttt{ESTAR}, \citealt{ESTAR92}), and $\omega_p$ is the plasma frequency.

The fraction of the total stopping power contributed by $dE/dx_{\rm ion,ex}$ will determine the fraction of energy that goes into excitations and ionizations. We assume the energy is then evenly split between excitations and ionizations to give the equation

\begin{equation} \label{eq:fion}
f_{\rm ion} = 0.5 \frac{dE}{dx}_{\rm ion,ex}/\left(\frac{dE}{dx}_{\rm elec} + \frac{dE}{dx}_{\rm ion,ex}\right)
\end{equation}

\noindent For non-ionized material, $f_{\rm ion}$ approaches 50\% when $dE/dx_{\rm ion,ex} \gg dE/dx_{\rm elec}$ while highly ionized material typically has a value of $f_{\rm ion}\sim3\%$.

\begin{figure*}[t]
    \centering
    \includegraphics[width=0.97\textwidth,trim={0cm 0.3cm 0cm 1cm},clip]{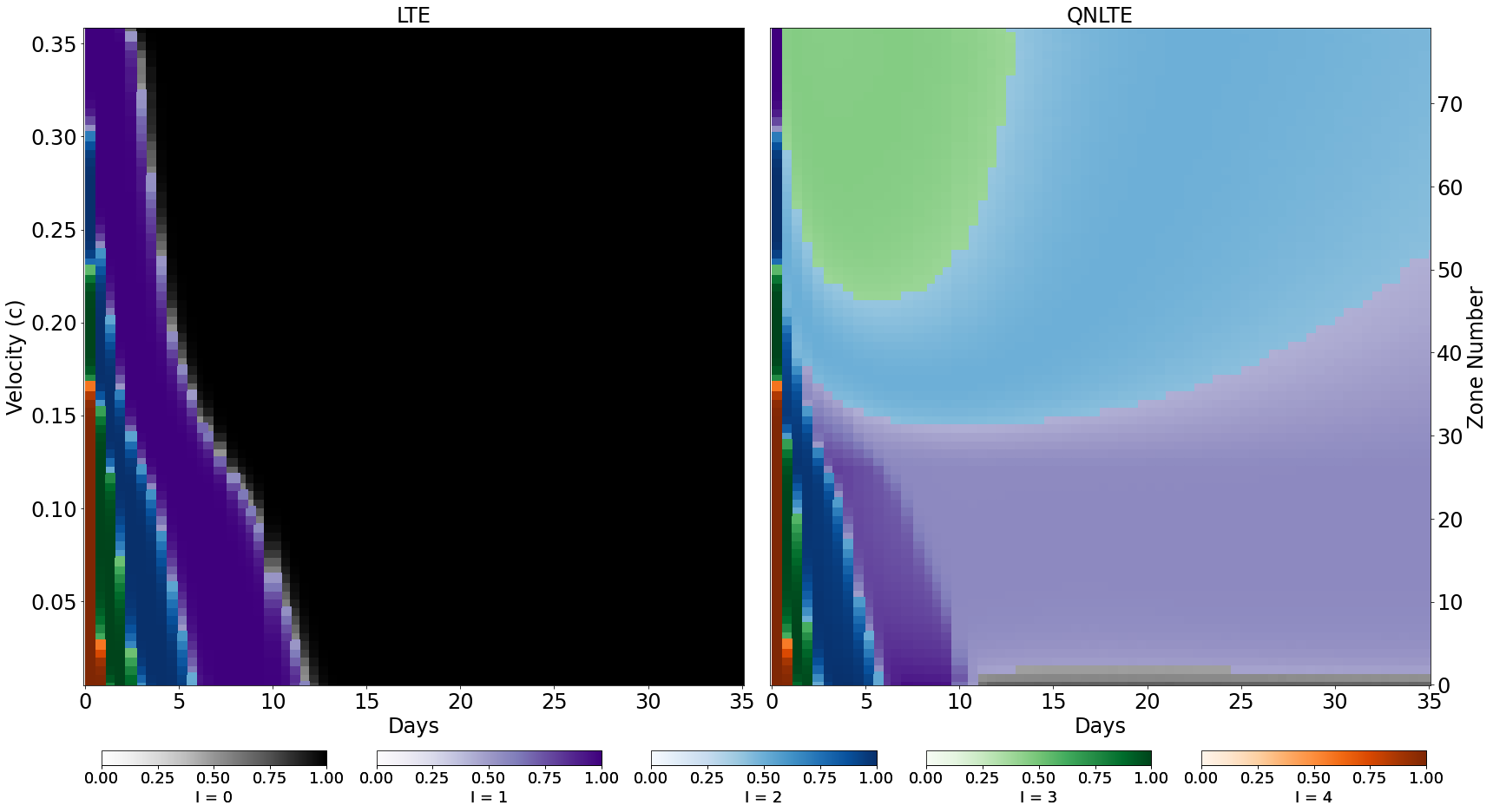}
    \caption{LTE (left) and QNLTE (right) ionization structure of Nd (Z=60) until 35 days post merger for the fiducial model of \Xlan\, = $10^{-2}$, $v_k$ = 0.1$c$, and $M$ = 0.03 M$_\odot$ ejecta. Color represents the dominant ionization species in a given zone at a given time, with intensity representing the fraction of that atom that is in the dominant ion state. Nd I, Nd II, Nd III, Nd IV, and Nd V are represented by black, purple, blue, green, and orange, respectively. In LTE, the ejecta at a given location are dominated by a single ionization species. By $\sim$ 13 days, the entirety of the ejecta are neutral. Conversely, when including radioactive ionization, the lowest density and coolest material in the outermost layers is the first to be significantly ionized up to the triply ionized state out to ten days. Unlike in LTE, the outermost layers are a ``blend" of higher ionization state species above the photosphere.}
    \label{Fig:NdIonStruct}
\end{figure*}

\begin{figure*}[h]
    \centering
    \includegraphics[width=0.98\textwidth,trim={1cm 0.8cm 0cm 2cm},clip]{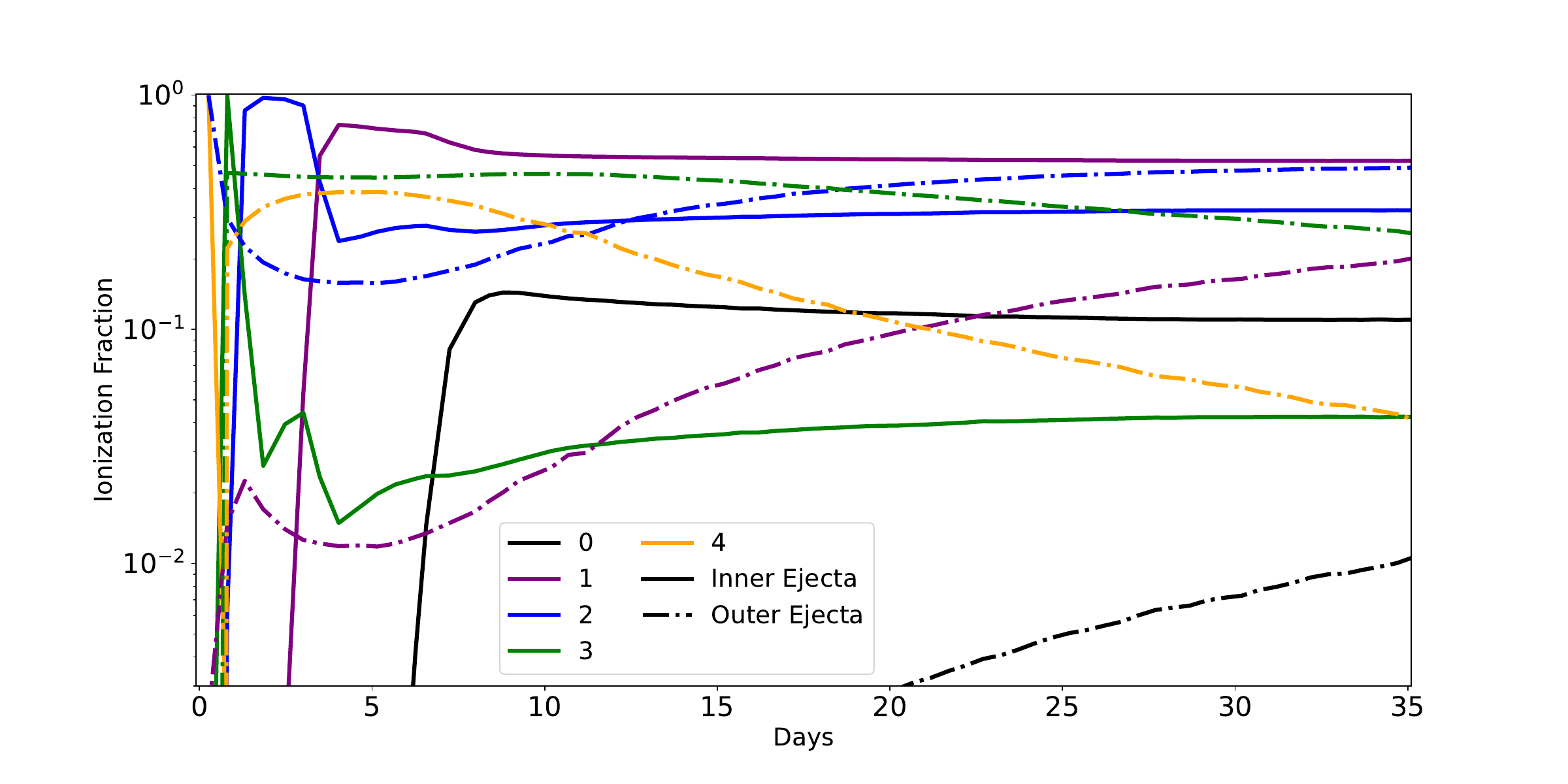}
    \caption{Ionization structure of Nd until 35 days post merger of the fiducial model with a representative inner ejecta zone (solid) and outer ejecta zone (dash-dotted). The color pattern is the same as Fig. \ref{Fig:NdIonStruct}. The inner ejecta eventually achieve a steady-state by $\sim 8$ days of $\sim 60\%$ singly ionized, $\sim25\%$ doubly ionized, and $\sim15\%$ neutral. The outer ejecta evolve more slowly but are more ionized than the inner ejecta due to the lower density.
    }
    \label{Fig:IonSliceNd}
\end{figure*}

\subsection{Atomic Data}

All models use atomic data generated by the Hebrew University Lawrence Livermore Atomic Code (\texttt{HULLAC}, \citealt{Bar-Shalom01}) for elements Z = 26-88, up to triply ionized species, in a self-consistent and systematic way for a large number of elements, as presented in \cite{Tanaka20}. For this work, we use elements Z = 31-70. All opacities employ the ``Sobolev expansion opacity" for binning the large number of lines from lanthanides. The ``Sobolev expansion opacity" uses the Sobolev approximation \citep{Sobolev60} for bound-bound transitions, which is applicable when the thermal line width of a given line is negligible compared to that of the expansion velocity. This is true for kilonova ejecta, as the expansion velocity is typically of the order 10$^3$ km/s while the thermal velocities are of the order of 1 km/s \citep{Kasen13}. The lines are then binned within the broader frequency bins of the simulation. We include effects from free-free and electron scattering opacities in each model as they can become more important at longer wavelengths ($\lambda \gtrsim {\rm few}\, \mu$m) where lanthanides no longer dominate as strongly. We do not consider bound-free absorption when calculating the opacity due to the negligible contributions of bound-free opacity compared to the bound-bound absorption from lanthanides (i.e., $\kappa_{bf, Nd} \ll  \kappa_{bb, Nd}$ until $\sim 10$ eV for kilonova-like conditions, \citealt{Fontes20}) and the wavelengths of many higher ionization state bound-free photons are outside the wavelength range simulated.

\subsection{$r$-process Heating} \label{Subsec:rproc}

Each zone in the simulation is assumed to include an $r$-process radioactivity. All zones use the same radioactive decay energy rate per unit mass given by:

\begin{equation} \label{Eq:heat}
    \dot{Q_r} = At^\alpha + B_1 e^{-t/\beta_1} + B_2 e^{-t/\beta_2}
\end{equation}

\noindent where A = 8.49$\times10^{9}$ erg g$^{-1}$ s$^{-1}$, $\alpha$ = -1.36, $B_1$ = 8.34$\times10^{9}$ erg g$^{-1}$ s$^{-1}$, $\beta_1$ = 3.63 d, $B_2$ = 8.86$\times10^{8}$ erg g$^{-1}$ s$^{-1}$, and $\beta_2$ = 10.8\,d as defined by the heating rate per unit mass for material of electron fraction $Y_e$ = 0.13, entropy per baryon of 32$k$, and expansion timescale 0.84 ms from \cite{Lippuner15}. We restrict our models to a single $r$-process heating rate to isolate the effects of altering the ionization state of the ejecta on the observables and limit other error sources (e.g., \citealt{Sarin24}). 

Only a fraction of the energy is thermalized due to escaping neutrinos, gamma rays, and ejecta-decay product interactions, which define the thermalization efficiency. We convolve the radioactive decay energy rate with the local thermalization prescription (see \citealt{Barnes16,Brethauer24} for more details) to determine the total energy each zone receives.

\section{Ionization Structure and Observables \label{Sec:IonStruct}}

\begin{figure*}
    \centering
    \includegraphics[width=0.89\textwidth,trim={2.5cm 1.6cm 3cm 2.7cm},clip]{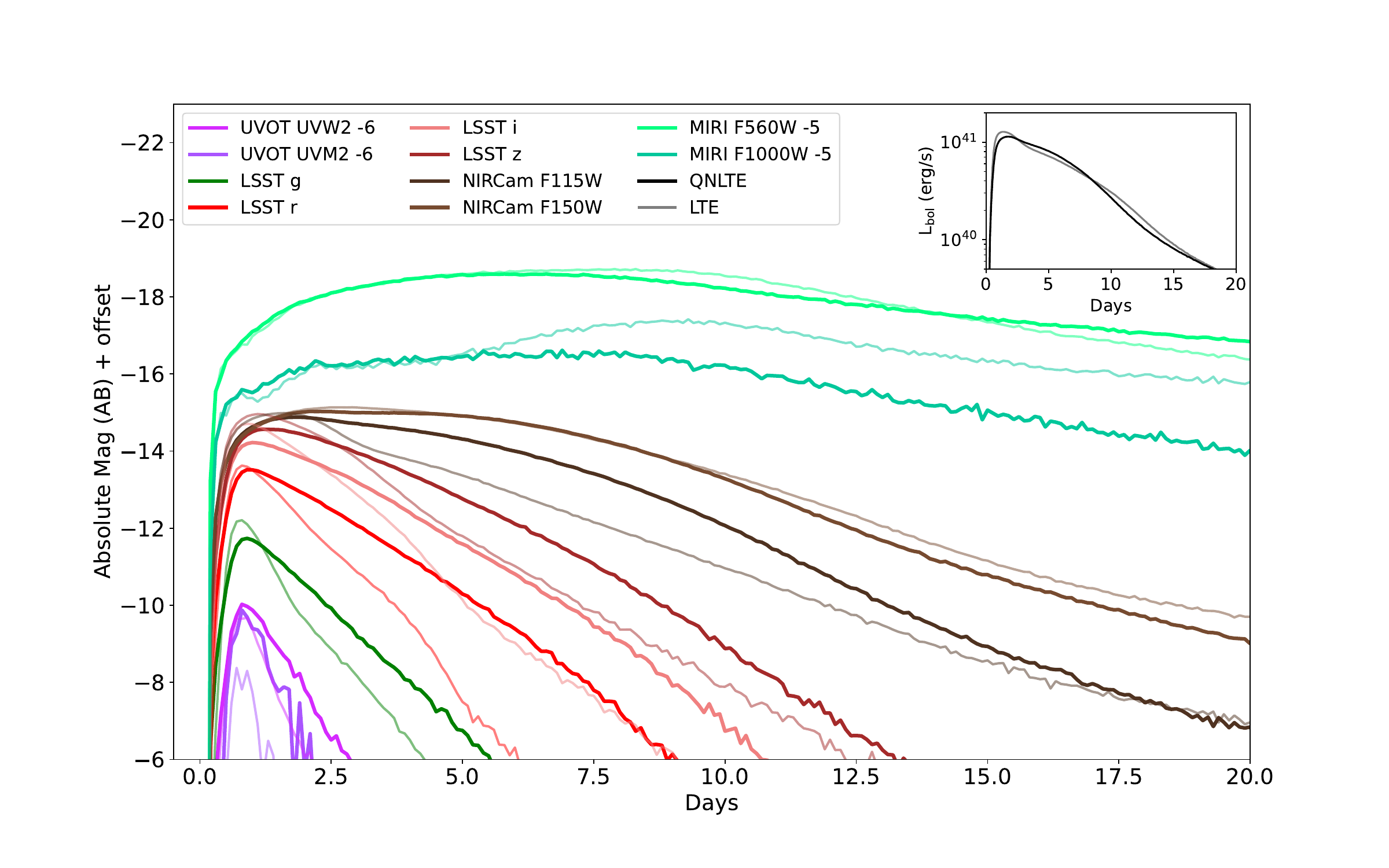}
    \caption{Absolute magnitude light curve of the fiducial ejecta model using QNLTE (dark) and LTE (light) in filters spanning from UV to MIR with bolometric luminosity inset. The light curves show dimmer peaks in each filter in QNLTE than LTE by less than a magnitude, but rapidly develop different decay rates, particularly at optical bands. The lower optical opacity in the QNLTE model prevents reprocessing of optical photons into longer wavelengths leading to longer diffusion timescales, resulting in generally brighter and longer lasting emission post peak.
    }
    \label{Fig:FiducialLC}
\end{figure*}
In this section, we explore how NLTE effects modify the ionization structure of kilonova ejecta. Specifically, we examine the ionization state of the representative lanthanide element Nd (Z = 60) for the fiducial model. As all other lanthanides have similar ionization energies, their ionization structure is nearly identical. 

In LTE models, the ionization structure of the ejecta is  determined by the Saha equation, which in turn depends on the temperature and ionization energy of the species. The left panel of Fig. \ref{Fig:NdIonStruct} shows the LTE evolution of the ionization fraction as a function of both time and space. In accordance with the Saha equation, each location in the ejecta at a given time is dominated by a single ionization state (except in the narrow temperature range where a species transitions from one ionization state to another). At early times when the gamma rays and high-energy electrons thermalize completely due to high densities trapping any radiation, diffusion produces a temperature gradient that monotonically decreases with radius. As the ejecta expand and cool due to adiabatic losses and radiative cooling, the ejecta decrease in ionization state beginning with the outermost material until it is completely neutral by $\sim 13$ days post merger.

The LTE and QNLTE models are similar at early times and in the inner layers, where the ejecta are the hottest and densest. At later times, however, radioactive ionization dramatically alters the ionization structure, especially in the low-density outer layers (right side of Fig. \ref{Fig:NdIonStruct}). Qualitatively, there are two distinct properties of ejecta with radioactive ionization: i) an ``inverted" ionization structure, with higher ionization at larger radii and ii) multiple ionization states located in the same physical region (which we call a blend of higher ionization states) throughout the low-density ejecta. 

To the first point, due to the radially decreasing density profile of the kilonova ejecta, the outer layers ($v >  v_t$) drop below the critical density, $n_{\rm nt}$, first, resulting in highly ionized species. In the inner layers of ejecta $(v <  v_t)$, the recombination rate is enhanced by the higher density and so species are less ionized. If the density is sufficiently high $(n > n_{\rm nt})$, LTE conditions hold and radioactive ionization does not alter the ionization state. Outside this LTE region, radioactive ionization generates a gradient of ionized species with the most highly ionized material in the most distant ejecta. 

To the second point, in the low-density regions of the ejecta (outermost at early times and most ejecta at late times), the ionization states are highly mixed without a single state dominating a given zone (Fig. \ref{Fig:IonSliceNd}). For example, after $\sim10$ days Nd in the inner power law component of ejecta is comprised of $\sim$ 60\% singly ionized material, with the remainder split between neutral and doubly ionized species. The material in the outer power law region is an evolving mixture of triply, doubly, and singly ionized species with each species between $\sim20\%$ and $\sim50\%$. Crucially, this provides an avenue for  lines from many different ionization species to appear in the spectrum at the same epoch, a behavior that is difficult to explain in 
LTE without fine-tuning, as suggested in AT\,2017gfo (e.g., \citealt{Hotokezaka23,Gillanders24,Sneppen24}). We explore this further in \S \ref{subsubsec:Lines}.

\begin{figure*}
    \centering
    \includegraphics[width=0.90\textwidth,trim={0cm 0.3cm 0cm 0.2cm},clip]{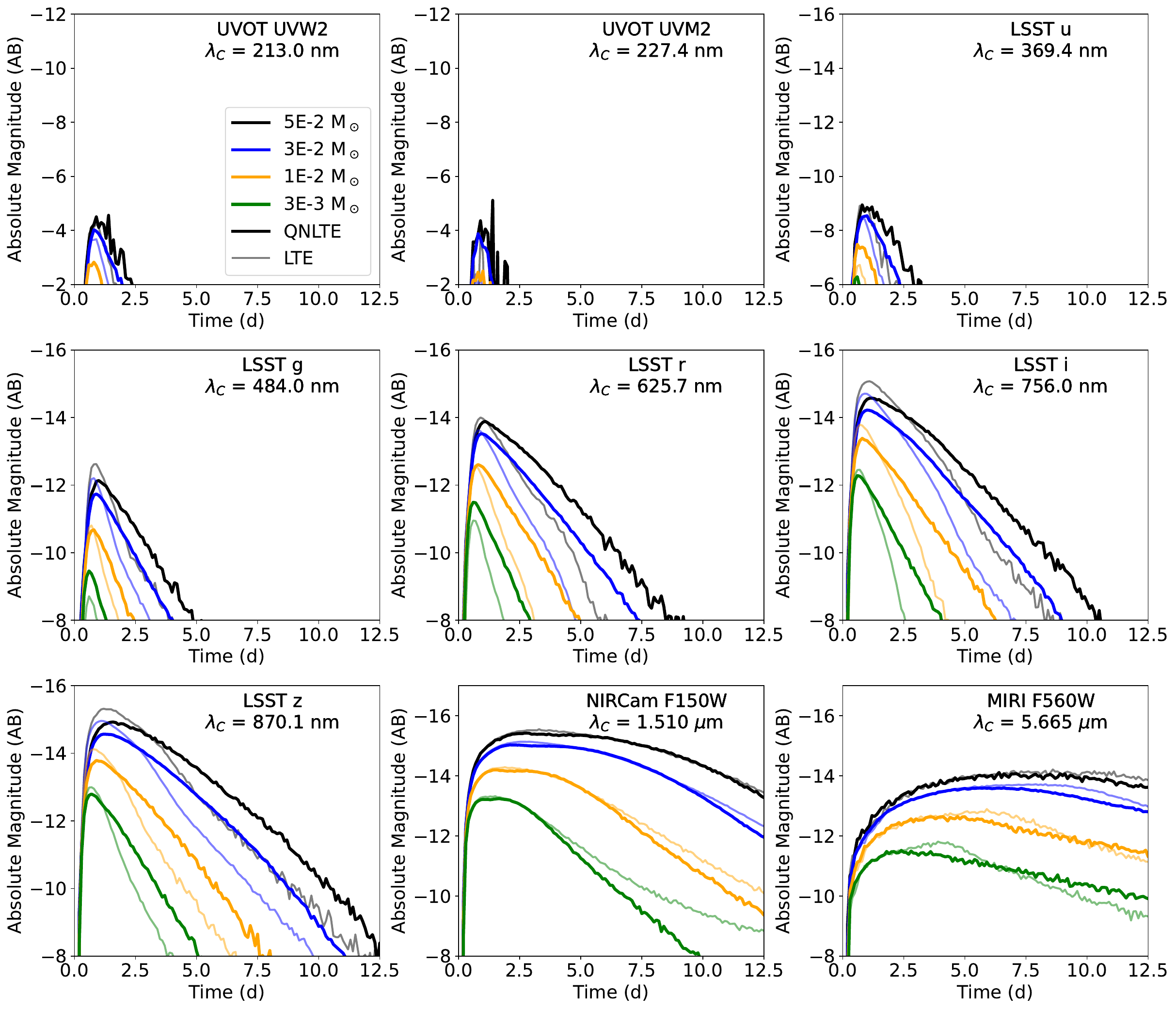}
    \caption{Broadband light curves of models with varying mass for QNLTE and LTE models. Critically, the QNLTE models differ by having significantly enhanced late-time optical emission and diminished NIR emission compared to the LTE models. LTE is a better approximation until later times for more massive models.}
    \label{Fig:MassLightCurve}
\end{figure*}

The inverted ionization structure and blend of ionization states in the ejecta alters the resulting UV to MIR light curve without dramatically changing the bolometric luminosity, presented in Fig. \ref{Fig:FiducialLC}. Generally, the peak of each light curve is dimmer by less than a magnitude and flatter when including QNLTE.  However, the decay rate of each light curve is substantially shallower, particularly at optical wavelengths, leading to longer lasting light curves and a brighter post-peak luminosity over the first week.  Lacking the higher line blanketing of neutral atoms, less optical emission may be reprocessed to longer wavelengths where the opacity is lower. Thus, photons are more trapped around peak, leading to longer diffusion times, a dimmer peak, a longer lasting light curve, and reduced emission at wavelengths beyond 6 $\mu$m. 

\begin{figure*}
    \centering
    \includegraphics[width=0.90\textwidth,trim={0cm 0.3cm 0cm 0.2cm},clip]{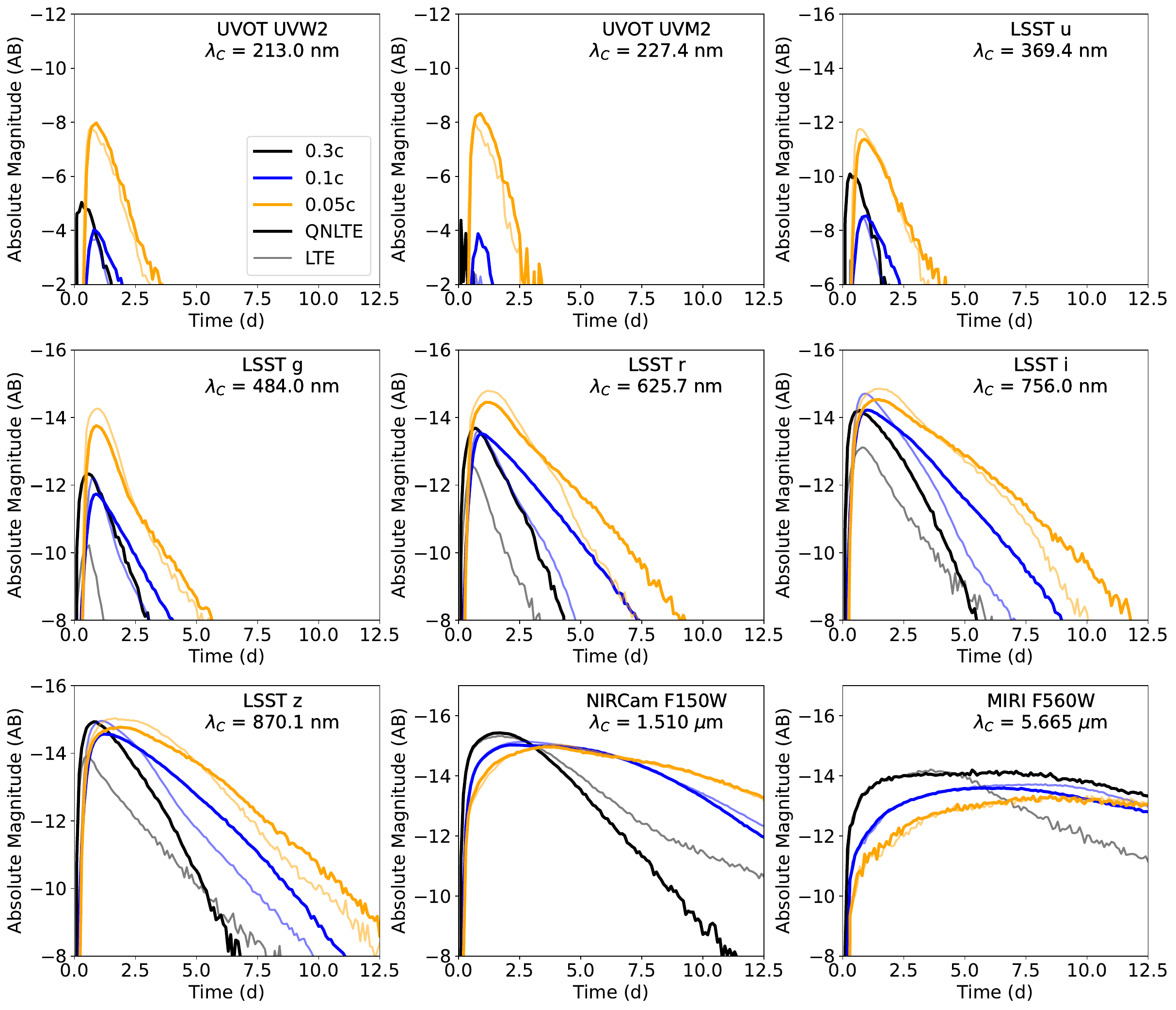}
    \caption{UV to IR light curves of kilonova models to demonstrate dependence of the broadband light curves on the characteristic velocity of the ejecta for QNLTE and LTE models. The QNLTE simulations are all consistently brighter at late times for bluer wavelengths, though the difference between QNLTE and LTE is not as prominent for the slowest moving ejecta. The increase in optical emission also corresponds to the decrease in NIR emission as less energy is reprocessed to longer wavelengths.}
    \label{Fig:VelLightCurve}
\end{figure*}

\begin{figure*}
    \centering
    \includegraphics[width=0.90\textwidth,trim={0cm 0.3cm 0cm 0.2cm},clip]{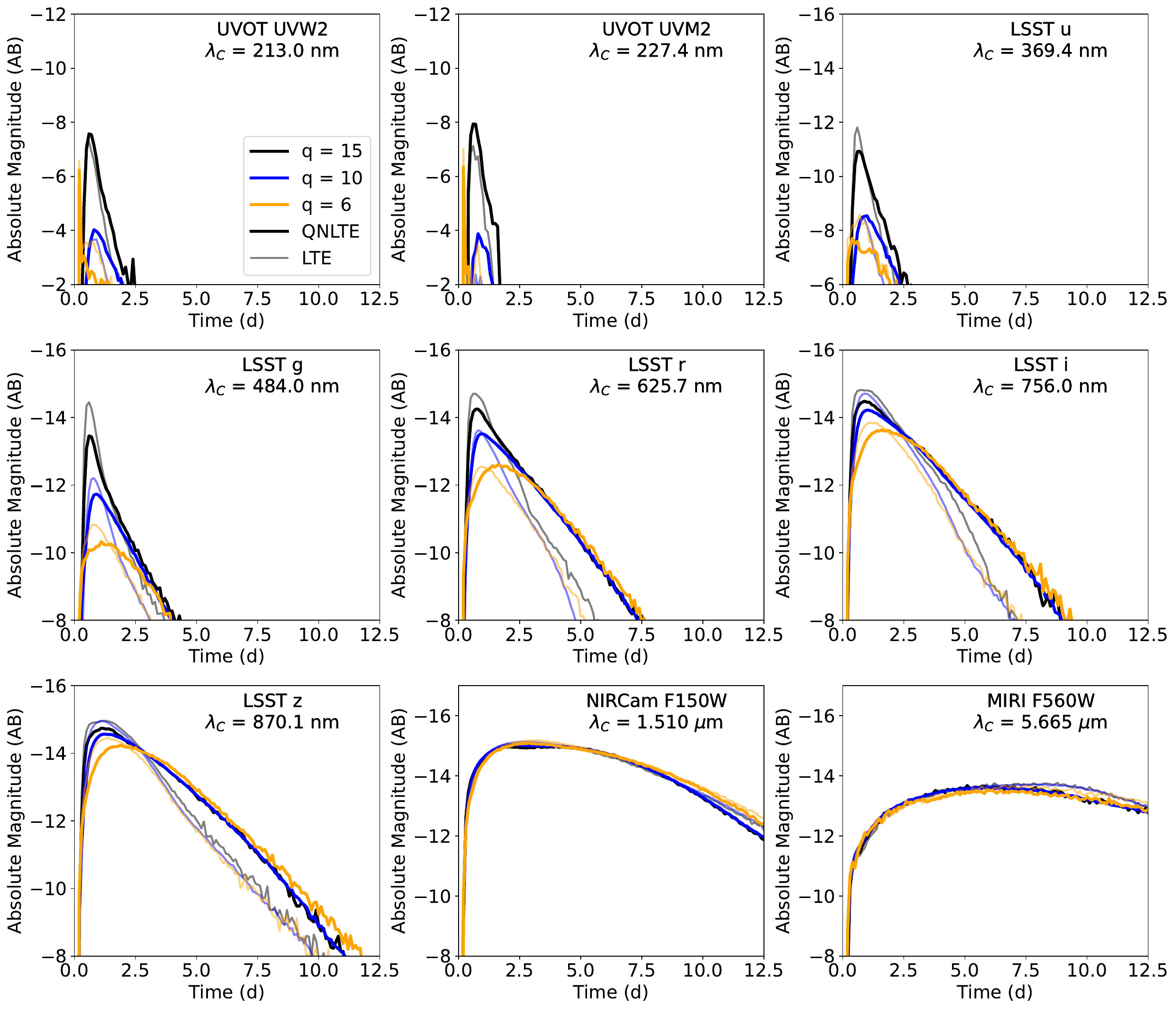}
    \caption{Broadband light curves of models with varying outer density profile power-law index, $q$. The peak of the light curve is most strongly affected by the density profile, with the effect becoming weaker at longer wavelengths. At IR wavelengths, the light curves become indistinguishable demonstrating that the IR is completely insensitive to the outer density profile of the ejecta.}
    \label{Fig:RhoLightCurve}
\end{figure*}

As the fiducial model is only one realization of kilonova ejecta, we now explore the impact of non-thermal ionization on a broader set of models. Figures \ref{Fig:MassLightCurve}, \ref{Fig:VelLightCurve}, \ref{Fig:RhoLightCurve}, and \ref{Fig:XlanLightCurve} show the UV to MIR light curves using both the QNLTE (dark) and LTE (light) treatments for models varying in velocity, mass, density profile, and \Xlan. 

\subsection{Dependence on Mass} \label{subsec:mass}

For models of higher mass, the ejecta remain hotter and denser longer, delaying the time that the ejecta have  $T < T_{\rm nt}$ and $n <n_{\rm nt}$ and indicating that an assumption of LTE is more valid at later times for more massive models (Fig. \ref{Fig:MassLightCurve}). Qualitatively, the ionization structure of the low and high-mass models are very similar to that of the 0.03 M$_\odot$ model in Fig. \ref{Fig:NdIonStruct}; the inner ejecta maintain a singly ionized state, there is a triply ionized ``bubble" of material within the outer ejecta component for the first $\sim2$ weeks caused by $t_n < t_{\rm \beta, th}$, and have LTE-like conditions in the densest ejecta at early times. However, the $M$ = 0.003 M$_\odot$ model deviates from LTE sooner due to achieving $n <n_{\rm nt}$ faster. For sufficiently low mass models like $M$ = 0.003 M$_\odot$, radioactive ionization dominates so strongly over recombination that of the inner region of the broken power law, the dominant ionization state is doubly ionized.

\subsection{Dependence on Characteristic Velocity} \label{subsec:vel}

At higher characteristic velocities, the density of the ejecta drops more quickly, resulting in higher ionization states at earlier times compared to slower counterparts. Similarly to mass, this correlates to the timescale that the ejecta deviate from LTE. For example, the bulk of the ejecta in the $v_k = 0.05c$ model remains in LTE until approximately 15 days post merger (as opposed to $\sim 1$ week in the 0.1c model). This manifests in Fig. \ref{Fig:VelLightCurve} with the faster QNLTE models deviating from LTE on shorter timescales, even pre-peak for UV and optical filters for the $v_k$ = 0.3$c$ model to the point that the UV light curves no longer monotonically decrease with higher velocity, though are still faint due to the high \Xlan.

\subsection{Dependence on Density Profile} \label{subsec:rho}

\begin{figure*}
    \centering
    \includegraphics[width=0.90\textwidth,trim={0cm 0.3cm 0cm 0.2cm},clip]{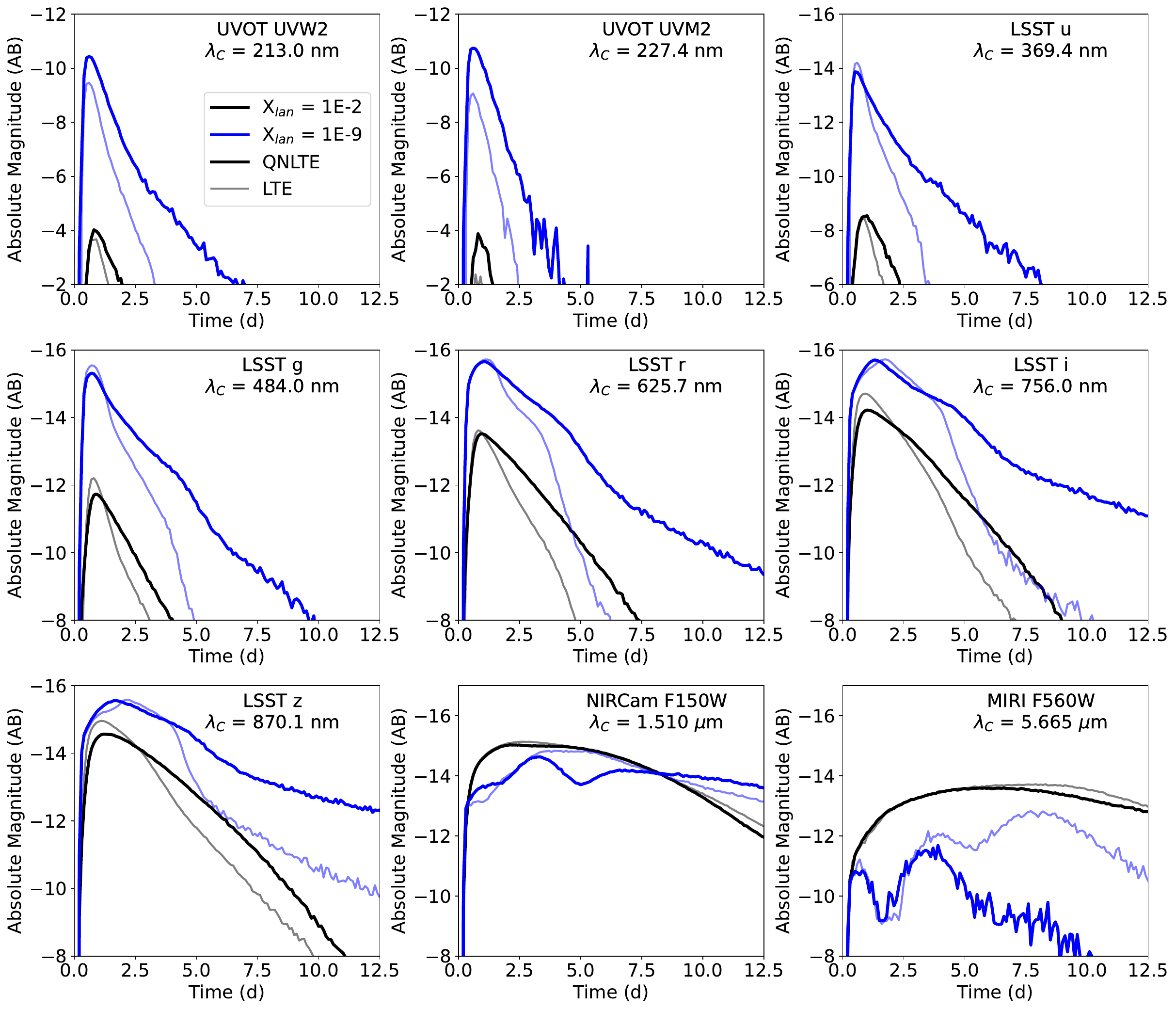}
    \caption{Dependence of the broadband light curves on the lanthanide content of the ejecta in QNLTE and LTE models. The QNLTE models exhibit a much shallower decay rate starting at $\sim4$ days, resulting in much longer lasting optical light curves that are more akin to what was seen in AT\,2017gfo. Even when lanthanides are not the dominant source of opacity, QNLTE effects significantly alter the ionization state and therefore the light curves.
    At later times, the low \Xlan\, model appears brighter at $\sim1.5$ $\mu$m, though we caution that this is the result of a single emission feature that may not be properly calibrated in the atomic dataset.
    }
    \label{Fig:XlanLightCurve}
\end{figure*}

\begin{figure*}
    \centering
    \includegraphics[width=0.98\textwidth,trim={1cm 0.8cm 0cm 2cm},clip]{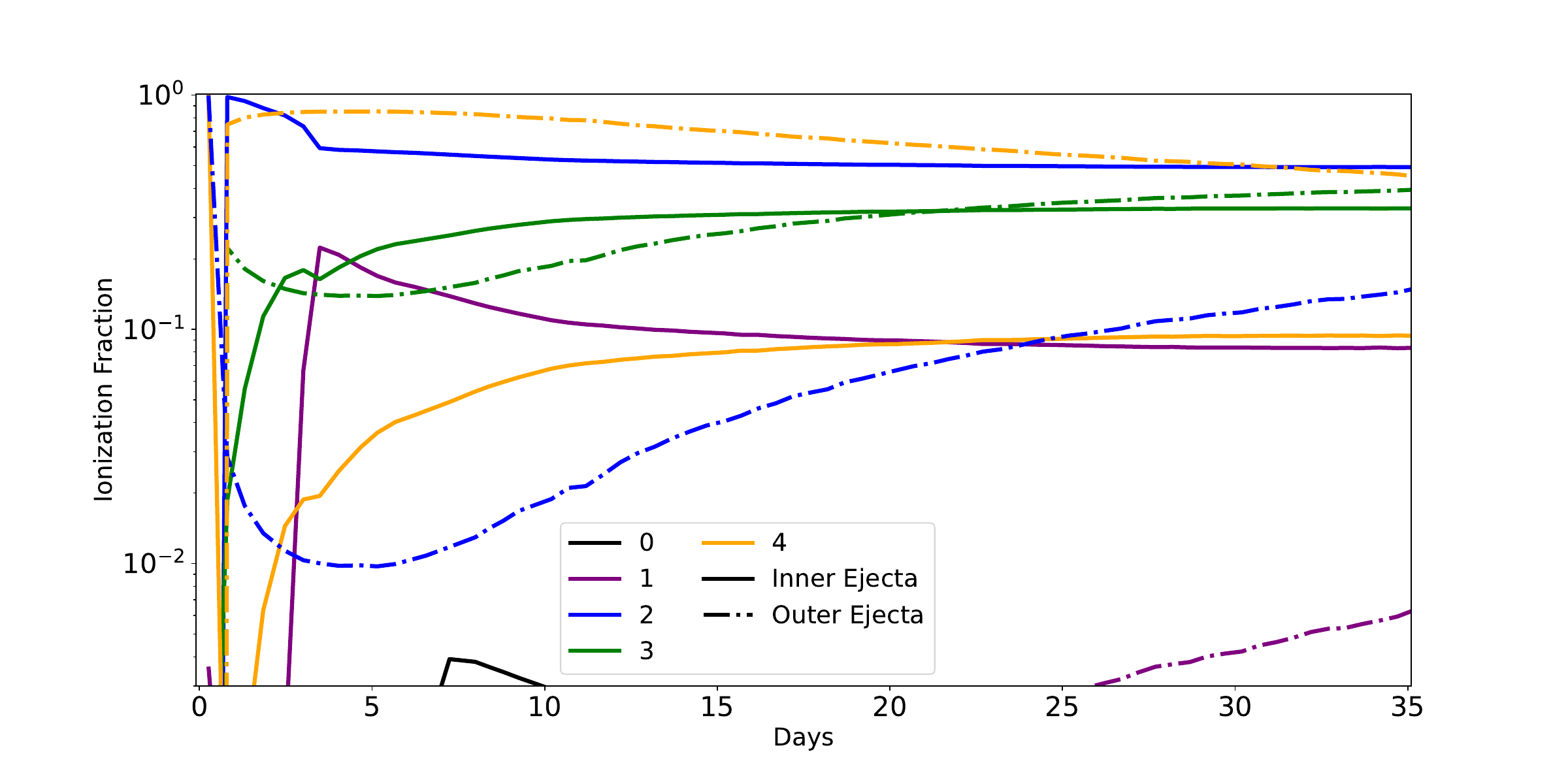}
    \caption{1D slice of the ionization structure of Sr (Z=38) until 35 days post merger for a low-lanthanide \Xlan = $10^{-9}$ model with a representative inner ejecta zone (solid) and outer ejecta zone (dash-dotted). The color scheme is the same as Fig. \ref{Fig:NdIonStruct}. The ionization state of the inner ejecta remains mostly constant beginning $\sim4$ days post merger with a mixture of singly, doubly, triply, and quadruply ionized species.}
    \label{Fig:IonSliceSr}
\end{figure*}

We vary the outer power law index of the density profile from the fiducial value of $q = 10$ to $q = 15$ and $6$ while keeping the inner power law index constant at $\delta = 1$. For a given total mass, a lower $q$ increases the fraction of mass in the outer layers. In our models, this results in a less ionized outer component due to the higher densities. The ionization state of the inner component is largely unaffected by changes to the outer power law, as the density in the inner layers only weakly depends on $q$.

The impact of the density profile on observables is strongest at the peak of the light curve, especially in bluer filters (Fig. \ref{Fig:RhoLightCurve}). The $u$ and $g$ light curves show a spread of 3 magnitudes at peak, and the sharpness of the peak increases with increasing $q$.

At $t \gtrsim$ few days, the LTE and QNLTE models converge at optical wavelengths as the photosphere passes through the outer power-law component, though the QNLTE models remain brighter than LTE by 1-2 magnitudes. Critically, at IR wavelengths the light curve appears entirely unaffected by the variation in density profile, indicating that the IR can serve as a more reliable measure of robust kilonova properties due to lower IR opacities and therefore an IR photosphere deeper in the ejecta.

Additionally, our results imply that the kilonova emission coming from broken power law (e.g., \citealt{Bulla23,Fryer24,Brethauer24}), single power law (e.g., \citealt{Bulla23,Fryer24,Banerjee25}), and constant (e.g., \citealt{Pognan22b}) density profiles will be affected differently by radioactive ionization. Ejecta with density profiles that are shallower than the broken power law used in this work, $q < 10$ or $\delta < 1$, will be more resistant to radioactive ionization, as a result of more time spent above $n_{\rm nt}$. Thus, the variation in density profiles can explain some of the range in the time of onset of strong NLTE effects at optical wavelengths compared to other works, from $\gtrsim3$ days to $\sim$ weeks \citep{Kasen13,Banerjee22,Pognan22b,Pognan22a}. 

\subsection{Dependence on Lanthanide Content}

The ionization structure of the ejecta for a given atom only has a minor dependence on the chemical composition of the surrounding material parametrized by \Xlan\, through $f_{\rm ion}$, as $\langle I \rangle$ in the Bethe formula will decrease with lower lanthanide fraction (Eq. \ref{Eq:bethe}). 
In low lanthanide content ejecta (\Xlan $\lesssim 10^{-4}$), the opacity will be set by the light $r$-process elements and the ionization state of elements like Sr (Z = 38) determine what spectral features are present. 

Figure \ref{Fig:IonSliceSr} shows the evolution of the ionization state for Sr in representative inner and outer layers of low \Xlan\,  ejecta. Dielectronic recombination is expected to be less important for non-lanthanides, leading to lower overall recombination rates  (see \S \ref{subsec:limits} for more details). As a result, the outermost layer are predominantly Sr V.  The atomic data from \cite{Tanaka20} do not include line data for the quadruply ionized stages, but at such low density, highly ionized regions are not expected to have significant effect on the opacity at optical wavelengths. Qualitatively, all light $r$-process elements show a similar overall ionization behavior as the lanthanide elements 
(Fig.~\ref{Fig:NdIonStruct}) such as the inverted ionization structure and the blend of higher ionization states. However, the light $r$-process elements settle into the approximately constant ionization state phase faster than the lanthanides. 

Both the inverted ionization structure and blend of ionization states affect the light curve at low \Xlan, as seen in Fig. \ref{Fig:XlanLightCurve}, which shows the light curves for models with \Xlan\, = $10^{-2}$ and $10^{-9}$. In UV and optical bands, both high and low \Xlan\, LTE models show a sudden steeper decay rate change around 4 days that is either shallower or non-existent in QNLTE models, at approximately the same time that the constant ionization state is achieved for lanthanides. The difference in light curves of the QNLTE and LTE models is more pronounced in the low \Xlan\, ejecta at late times, growing to as much as 5 magnitudes at optical wavelengths. Without as much optical light reprocessing, the MIR of the QNLTE models tends to be fainter.

\begin{figure*}[t]
    \centering
    \includegraphics[width=0.85\textwidth,trim={0cm 0.26cm 0cm 0.2cm},clip]{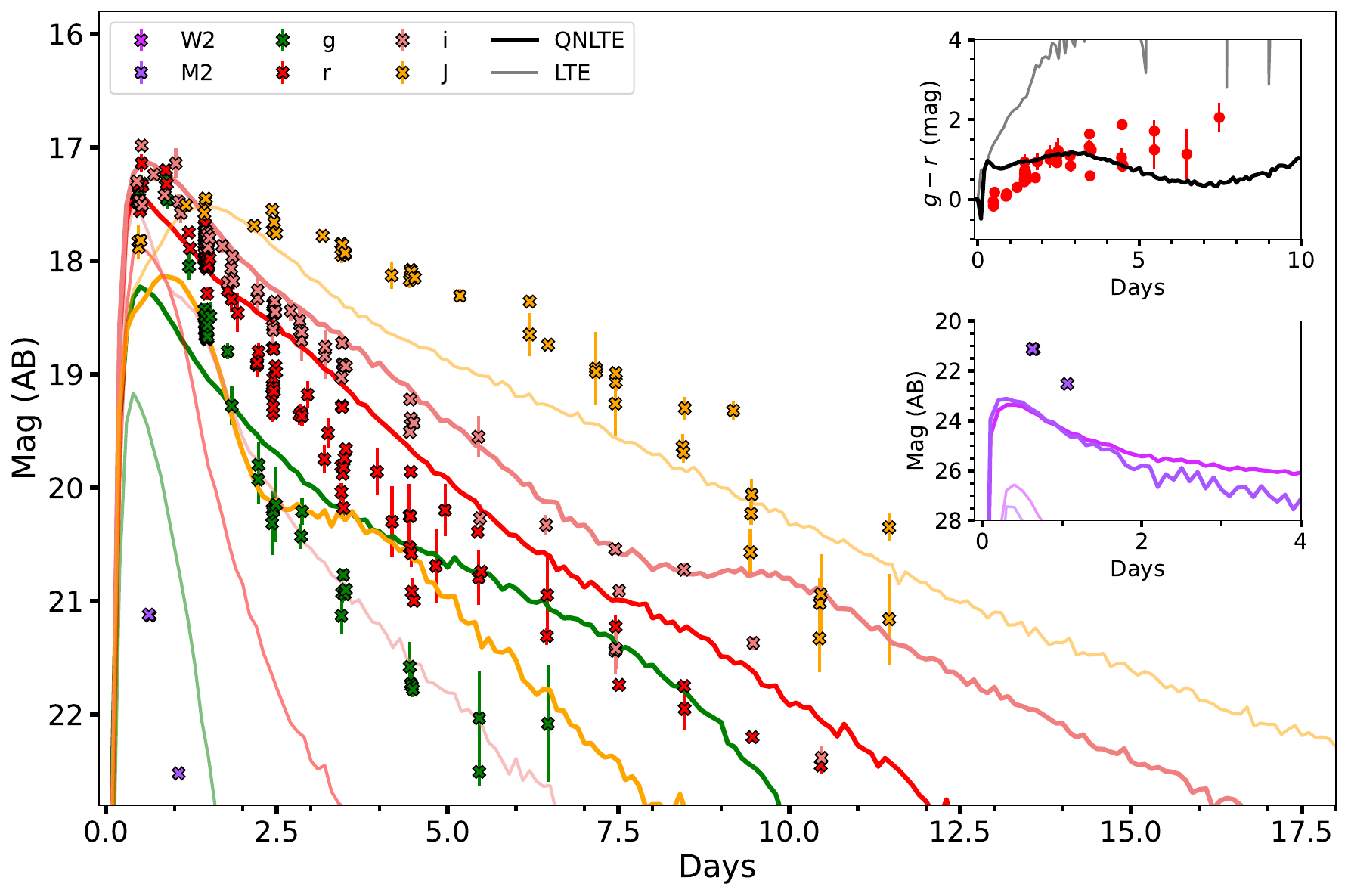}
    \caption{$g,r$, $i$, snd $J$ band light curves of AT\,2017gfo (compiled by \citealt{Villar17}) to the QNLTE and LTE models with \Xlan\,$= 10^{-9}$, $M$ = 0.03 M$_\odot$, and $v_k$ = 0.3$c$. With radioactive ionization included, the optical light curves maintain a significantly higher flux and the NIR light curve is significantly fainter compared to the LTE counterparts, requiring the inclusion of a red component to match observations of AT\,2017gfo. 
    \emph{Upper Inset:} $g - r$ color for the QNLTE and LTE models shown in left panel compared to AT\,2017gfo. The LTE model rapidly reddens within $\sim2$ days to $g-r\sim4$ mags, while the QNLTE model maintains a $g-r$ of 1 to 2 mags, in closer similarity to AT\,2017gfo. At later times, the red component will add additional red flux and could compensate for the lanthanide-free model being too blue.
    \emph{Lower Inset:} Zoom-in on the UV detections of AT\,2017gfo. The QNLTE model is too faint $\sim 2$ mags, but brighter than the LTE model by $\sim 3$ mags.}
    \label{Fig:17gfoComp}
\end{figure*}

\begin{figure*}[t]
    \centering
    \includegraphics[width=0.80\textwidth,trim={2.2cm 4.4cm 3cm 5.3cm},clip]{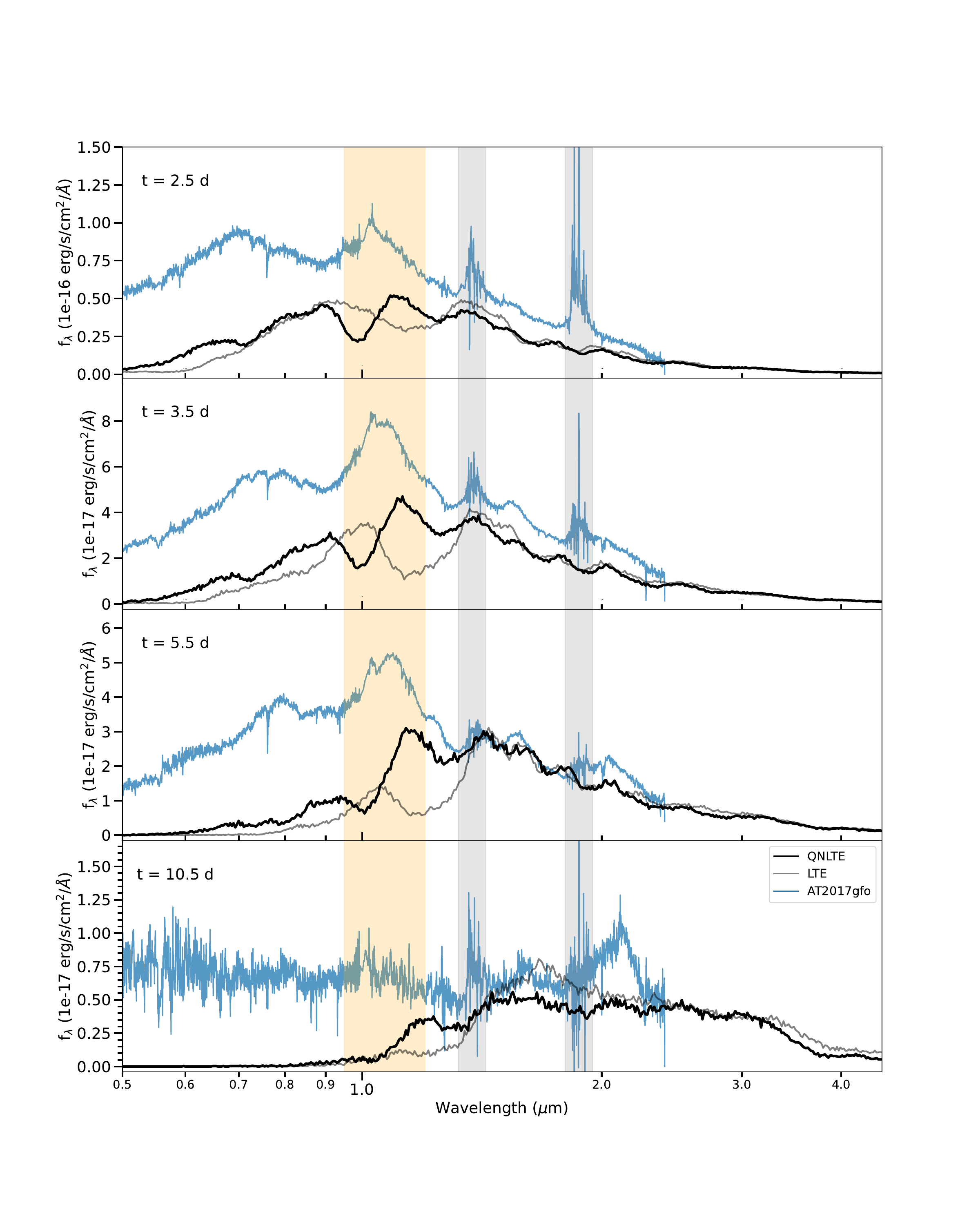}
    \caption{
    Spectral sequence of the fiducial model at a distance of 40 Mpc using QNLTE and LTE models. X-shooter data of AT\,2017gfo \citep{Pian17,Smartt17} are superimposed in blue. We emphasize that the fiducial model is not a fit to AT\,2017gfo, but the QNLTE model is an improvement towards matching spectral features such as the early-time peak around 1.05 $\mu$m, where the LTE models had a significant absorption feature (highlighted orange region). Regions with high telluric absorption are shaded gray.}
    \label{Fig:SpecSeq}
\end{figure*}

\subsection{Application to AT\,2017gfo} \label{Subsec:gfo}

As we have shown previously, LTE simulations produce fainter optical light curves with faster decay rates than what was observed with AT\,2017gfo. However, when including the effects of radioactive ionization these deficiencies are significantly reduced. Motivated by the fact that models with high lanthanide abundances cannot reproduce the early blue colors of AT\,2017gfo and that past work proposed a lanthanide-poor blue kilonova component as the dominant source of emission at early times (e.g., \citealt{Kasen17}), we present a comparison between a \Xlan\,$= 10^{-9}$, $M$ = 0.03 M$_\odot$, and $v_k$ = 0.3$c$ model and observed data in Fig. \ref{Fig:17gfoComp}. These parameters were chosen to approximately match the optical peak magnitudes of AT\,2017gfo. Despite not being a fit to AT\,2017gfo, the lanthanide-free QNLTE model is able to reproduce many of the observational features that the LTE model cannot: i) the emission is significantly bluer ($g-r \sim 1$ mag for the QNLTE model while $g-r \gtrsim 2$ mag for the LTE model); ii) the emission is brighter across all optical bands; and iii) the optical light curves maintain a slower decay rate out to late times resulting in detectable emission. Furthermore, the lanthanide-free component no longer produces significant NIR emission that is comparable to observations of AT\,2017gfo. Thus, QNLTE modeling requires the existence of another kilonova emission component, likely the lanthanide-rich ejecta from the merger. Proper fitting using multi-component and multi-dimensional radiative transfer simulations is beyond the scope of the current paper, and will refine estimations of the ejecta from AT\,2017gfo further. Specifically, the model presented in Fig. \ref{Fig:17gfoComp} is too bright in $r$ band at $t < 5$ days and in $g$ band at $t > 3$ days, which could be reconciled with a smaller ejecta mass, interactions between ejecta components, or viewing angle effects. Additionally, while our $v_k$ = 0.3$c$ and lanthanide-poor model is brighter in the UVOT $W2$ and $M2$ bands than LTE by 3 mags, it is still insufficient to explain the UV data of AT\,2017gfo. Models with more mass, steeper density profiles, or another energy source could help to explain the UV observations.

Accounting for non-thermal ionization is particularly important for understanding the origins of the blue kilonova component (\Xlan\,$\lesssim 10^{-4}$), as modeling points towards high ejecta velocities which rapidly have densities below $n_{\rm nt}$. The origin of the blue component has remained a topic of much debate \citep{Arcavi18}. Explanations range from shock-heated ejecta \citep{Oechslin07,Sekiguchi16,Piro18}, to post-merger disk ejecta driven by a combination of magnetic fields and neutrinos \citep{Metzger&Fernandez14,Metzger18,Miller19,Mosta20,Ciolfi20,Curtis24}, or spiral density waves from the remnant and accretion disk \citep{Nedora19}, which can be disentangled by reliable measurements of the ejecta properties. 


Beyond the light curves, the spectra of the QNLTE models reveal improvements to matching broad features of AT\,2017gfo over LTE models (Fig. \ref{Fig:SpecSeq}). We caution that the precise locations of features are likely to be inaccurate due to our use of uncalibrated line data, but the overall change in spectral shape from LTE to QNLTE models is robust. This is due to the density of lines from the lanthanides at optical wavelengths, where the statistical properties of the opacity remain largely consistent regardless of the precise location of any individual line. The lack of calibrated data is a problem across many kilonova models, presenting a tradeoff of either utilizing the limited calibrated atomic data for the strongest transitions only or using uncalibrated theoretical atomic data to capture the statistical properties of lanthanide opacities. This tradeoff emphasizes the need for additional experimental data.

The region around 1 $\mu$m is where the most significant changes from LTE to QNLTE occur. Most notably, the peak at $\sim1.05$ $\mu$m of AT\,2017gfo is more consistent with the QNLTE model (though the QNLTE model is slightly more redshifted), where the LTE model would predict a large absorption feature within the first 5 days. Additionally, the feature around 0.8 $\mu$m is fading faster in LTE than QNLTE models, resulting in the sharper and higher peak in the LTE $z$ band light curve in Fig. \ref{Fig:FiducialLC} at $t \lesssim 2.5$ days. As time progresses, at $\sim$ 0.6 $\mu$m the emission is stronger in the QNLTE model driving the longer-lasting light curve in $r$ band. 

\section{Discussion} \label{Sec:Discussion}

\subsection{Dynamical Ejecta}
Our analysis of the effects of radioactive ionization suggests that dynamical ejecta, due to their low mass and high velocities (e.g., \citealt{Bauswein13,Hotokezaka13,Sekiguchi16,Ciolfi17,Radice18}), could be ionized as the density rapidly drops below $n_{\rm nt}$ if the thermalization efficiency remains sufficiently high. This suggests that while the lanthanide-rich equatorial dynamical ejecta may play an important role in absorbing and reprocessing radiation from the blue kilonova component (e.g., \citealt{Kasen15, Kawaguchi18}), the effect may not be as strong as previously thought given the increased ionization state.  Further investigations into the effects of QNLTE on the dynamical ejecta and the interplay between low electron fraction heating rates and low thermalization efficiency could illuminate the observable signatures of the various mass ejection mechanisms.

\subsection{Extracting Ejecta Properties}

One consequence of the reduction in opacity due to radioactive ionization is that the kilonova is brighter optically post-peak and at peak for faster velocity models. Modeling of AT\,2017gfo (and other kilonovae, though with much higher uncertainties) tends to yield higher masses ($\sim$ few $\times 10^{-2}$ M$_\odot$, \citealt{Kitamura25,Rastinejad25}) than current GRMHD simulations would suggest are ejected from the merger system ($\lesssim 10^{-2}$ M$_\odot$ for dynamical ejecta, e.g., \citealt{Siegel19} and references therein). 

Including radioactive ionization effects helps to alleviate this tension, as more blue optical emission emerges from a QNLTE kilonova model than an LTE kilonova model with the same ejecta properties, especially for ejecta with higher velocity. Thus, for a given optical light curve, less mass is required to power the blue emission, which could result in either an overall reduction in mass or a shift in mass to the red component to conserve the bolometric luminosity. We provide a rough proxy for the difference in mass estimates by matching the peak brightness of a QNLTE model to a LTE model. In the case of a $M$ = 0.03 M$_\odot$, $v_k$ = 0.3$c$, and \Xlan\,= 10$^{-2}$ model, the peaks of the $r, i,$ and $z$ bands are equivalent to an LTE model of $M$ = 0.1 M$_\odot$ with the same $v_k$ and \Xlan\, (though the $g$ band of the LTE model is still too faint by $\sim$ 1 mag). Considering the same model but with \Xlan\,= 10$^{-9}$ yields less extreme mass reduction, requiring only $M$ = 0.05 M$_\odot$ in LTE to match the peak $g, r, i$, and $z$ bands (though the LTE model fades too quickly at times later than $\sim 2$ days). Thus, when considering only optical emission (as could be the case in a serendipitous discovery of a kilonova in archival data), the LTE models require a factor of $\sim 3$ and $\sim1.5$ more mass to fit a given optical light curve peak, respectively. 

For the low-lanthanide component, we consider the $\sim1.5$ factor more mass to be an upper limit, as the true difference between QNLTE and LTE model mass estimates may be smaller as the proper $r$-process heating rate will be lower than the function we assume (Equation \ref{Eq:heat}), resulting in less radioactive ionization and an ionization structure potentially closer to LTE. The NIR emission is fortunately distinguishable at low \Xlan \,(and even more so at high \Xlan) beginning around 4 days post merger. Similarly, the bolometric luminosity is distinguishable between models of different masses, which highlights the importance of obtaining IR observations to capture the full spectral energy distribution to constrain the mass of the ejecta. Additionally, since the fastest ejecta velocities have the highest discrepancy between QNLTE and LTE, the reduction in mass estimates will be smaller for slower ejecta velocities. Effectively, the difference in mass estimates will be highest for high lanthanide content and high velocity ejecta, which is most similar to the equatorial dynamical ejecta.

\subsection{Implications for Line Identification}
\label{subsubsec:Lines}
One hurdle to confident line identification in kilonova spectra is connecting the physical properties of the ejecta with the ionization states present across all species and at different epochs. For example, the identification of Sr II features out to late times \citep{Watson19,Sneppen23a,Gillanders24, Sneppen24} is problematic in the context of purely LTE ejecta, which are expected to be largely neutral at these phases. Similarly, highly ionized species like Te III, Se III, or W III have been associated to features in AT\,2017gfo \citep{Hotokezaka22,Hotokezaka23} or the kilonova that accompanied GRB 230307A \citep{Levan24}. The inclusion of radioactive ionization makes it possible for such high-ionization species to exist at late times. Additionally, there is a narrow window of temperature in LTE where Sr II or Y II can coexist with other ion species suspected to produce spectral features in the same epoch such as Ce III or La III due to similar ionization energies \citep{Sneppen24}. However, with the blended ionization structure that radioactive ionization creates, there is no longer a need to fine-tune the ejecta temperature. The identification of different ionization states of the same species at the same epoch would be significant evidence for the blended structure, and we encourage follow-up analysis of spectral features in AT\,2017gfo under this context.

\subsection{Modeling Uncertainties}

\label{subsec:limits}

Here, we briefly discuss the list of modeling uncertainties that are intrinsic to our models and their potential impacts on the observables, ordered approximately from most impactful to least.

\begin{itemize}

    \item \textbf{One $r$-process Heating Rate} - We use a single $r$-process heating rate function based on low electron fraction ($Y_e \lesssim 0.2$) ejecta. A more self-consistent approach would be to vary the $r$-process heating rate as a function of ejecta parameters such as $Y_e$, which would allow for more accurate mass estimates \citep{Sarin24}. However, the uncertainty in the $r$-process heating rate itself can vary by factors of a few depending on the nuclear mass model \citep{Barnes21}, comparable with the variations from different electron fraction values. In general, more accurate heating rates would likely result in the blue component becoming fainter due to the lower $r$-process heating rate from higher electron fraction material, and could change the shape and color of the light curves, affecting low-lanthanide models like those presented in Fig. \ref{Fig:17gfoComp}. As our fiducial model has a high lanthanide content and is therefore more consistent with the low electron fraction heating rate, we expect our results to not be severely impacted. 

    \item \textbf{Fraction of Beta Decay Energy in High-Energy Electrons} - \cite{Barnes16} and \cite{Wollaeger18} found that approximately 20\% of the energy from the $r$-process heating is carried by high-energy electrons. The precise fraction does vary with time at around $\pm3\%$ \citep{Wollaeger18}, and the choice of nuclear mass model can further vary the fraction by factors of a few \citep{Barnes21}. Additionally, if alpha particles become a significant fraction of the $r$-process heating rate, then alpha particles can efficiently thermalize via ionizations and excitations and would enhance the ionization rate. The combined effects of the uncertainty in the contributions from alpha particles and the fraction of energy emerging as high-energy electrons would scale the radioactive ionization rate and bolometric luminosity. The changes to the ionization rate would manifest as inverse changes to the recombination rate, as presented in Appendix \ref{App:Recomb}, while the kilonova would be brighter (dimmer) bolometrically due to the proportion of energy escaping as neutrinos or gamma rays decreasing (increasing).

    \item \textbf{Uncertainties in Early Blue Emission} - 
    As demonstrated in section \ref{subsec:rho}, the density profile of the ejecta influences the light curve shape at and around peak with steeper density profiles resulting in sharper peaks at UV and optical wavelengths. Furthermore, there is a confluence of factors that can produce or alter early time blue emission---shock interaction \citep{Piro18}, engine-driven energy injection \citep{Ai25}, density profiles \citep{Fryer24}, atomic structure of highly ionized lanthanides \citep{Banerjee24}, structured jets \citep{Kasliwal17,Gottlieb18a}, and variations in early-time $r$-process heating rates \citep{Sarin24}---that we do not include and make isolating any individual effect from observations of pre-peak light curves challenging. 

    \item \textbf{Spherically Symmetric Ejecta} - 1D simulations are able to capture the effects we examine in this work in a computationally efficient manner. Asymmetries and viewing angle effects have been shown to alter the light curve and therefore the derived ejecta properties (e.g., \citealt{Kawaguchi18,Wollaeger18,Korobkin21,Shingles23}). Moving to multi-D simulations would allow us to further refine our models and include effects like inter-component interactions, which could potentially solve the issue of models being too bright in the IR at early times ($t \lesssim 5$ days) seen in \cite{Brethauer24}. While our models are closer to the IR observations of AT\,2017gfo at $t \gtrsim 5$ days due to the decrease in optical/UV reprocessing from smaller opacities, the lack of change in NIR emission prior to peak suggests that QNLTE modeling cannot solve the overprediction of NIR flux at early times alone. The excess of NIR flux compared to AT\,2017gfo may be a result of neglecting multi-D effects, as the blackbody-like emission will be proportional to the projected surface area of the ejecta. Our implementation of QNLTE effects enables more rapid multi-D calculations that could efficiently investigate if the NIR excess at early times can be resolved with multi-D effects or alternative ejecta properties.
    
    \item \textbf{LTE Level Populations} - The same high-energy electrons that cause enhanced ionization states in the ejecta should also lead to changes in the level populations through non-thermal excitation. However, due to the high computational cost of calculating the full NLTE rate equations for all energy levels of lanthanides, we   have assumed LTE level populations. \cite{Pognan22b} show that the opacity significantly deviates beginning $t \gtrsim 5$ days as a result of NLTE level populations, suggesting that LTE populations are sufficiently accurate at earlier times. \cite{Pognan22b} also demonstrate that NLTE excitations reduce the overall opacity, which will have similar effects on the light curve to the radioactive ionization presented here. By focusing on the NLTE ionization state instead of the NLTE level populations, we capture the significant changes to opacity without increasing computational costs dramatically.
    
    \item \textbf{Composition} - The composition of our ejecta is based on solar and meteoric $r$-process abundances (with the $s$-process contributions removed) due to the universality of $r$-process abundances. This means that the \Xlan\, values we use are not correlated directly to a nucleosynthesis calculation. A composition taken directly from nucleosynthesis calculations would be more self-consistent, but there are factors of a few uncertainty in lanthanide abundances as a result of variable nuclear models \citep{Barnes21}. Furthermore, the line-blanketing in the optical produced by lanthanides as a whole may make the spectral shape insensitive to the precise abundances. Abundance patterns can alter the spectral features, particularly with species like Nd and Ce \citep{Kasen17}, but the broad color remains largely consistent.

    \item \textbf{Partition of Beta Particle Energy between Excitation and Ionization} - We define the fraction of energy that thermalizes via ionization from a high-energy electron produced by beta decay of $r$-process elements as $f_{\rm ion}$. $f_{\rm ion}$ is based on the Bethe formula (Eq. \ref{Eq:bethe}), which describes the energy that a charged particle loses to ionizations and excitations. We assume, simplistically, that the energy losses are evenly split between ionization and excitation, which may not always be true as the partitioning depends on the relative cross-sections to ionization and excitation. If a smaller fraction of energy goes into ionizing atoms the effect would be the same as an increase in recombination rate when radioactive ionization and radiative recombination dominate the rate equations, which we explore in Appendix \ref{App:Recomb} and find an overall decrease in blue light light curves with subtle changes increases to IR light curves. 
    
\end{itemize}

\section{Conclusions} \label{Sec:Conc}
Radioactive ionization due to energetic beta decay electrons from freshly synthesized $r$-process material plays a prominent role in determining the ionization structure of kilonovae, and therefore the opacity, light curves, and spectra. We have shown that including the effects of radioactive ionization and dynamically calculating the fraction of energy that ionizes the ejecta from beta decays ($f_{\rm ion}$) significantly increases the ionization state of optically thin ejecta near and above the photosphere, resulting in a lower opacity, inverted ionization structure (ionization state increases farther out in the ejecta), and coexistence of ionization states in the same physical region. These ionization structures are fundamentally due to the ejecta dropping below a critical density and critical temperature where radioactive ionization will significantly impact the ionization state.

Observationally, QNLTE models have a slower decay rate of optical light curves, bluer colors near and after peak, and a light curve shape that more closely matches AT\,2017gfo due to lower optical opacities, where previous LTE wavelength-dependent opacity models faced significant challenges doing so. As a consequence of the reduction in optical opacity, more optical light escapes, preventing photons from being reprocessed into IR (especially in high velocity components), and resulting in significantly fainter IR light curves that require a red component. Additionally, including QNLTE effects helps to alleviate some of the tensions in kilonova modeling. Specifically, i) compared to LTE models, QNLTE models of fast ($\sim 0.3c$) ejecta require a smaller ejecta mass to produce an equivalently bright optical light curve, bringing mass estimates closer to the smaller masses of dynamical ejecta predicted by GRMHD simulations; ii) the potential identification of highly ionized species at late phases, when LTE simulations predict only neutral atoms, can be explained by radioactive ionization; and iii) the ejecta affected by non-thermal ionization are a mixture of ionization states rather than one dominant state, eliminating the need for fine-tuning the temperature and density parameters to allow spectral features from a range of ionization states to coexist. 

The importance of radioactive ionization at low densities and temperatures emphasizes the necessity of including these effects for simulating dynamical ejecta. The dynamical ejecta are expected to have a smaller mass than the secular ejecta while having more than double the characteristic velocity, which could result in significant departures from LTE. In particular, QNLTE has the potential to strongly affect the equatorial dynamical ejecta and reduce reprocessing of photons from the blue kilonova component.
Further investigations on the interactions of QNLTE dynamical ejecta with photons from the blue kilonova component can illuminate the influence of dynamical ejecta on spectral features and make progress towards concretely identifying mass ejection mechanisms of BNS mergers. While full multi-level NLTE calculations may be computationally infeasible in multi-dimensional radiative transfer simulations, QNLTE captures some of the key NLTE effects without the high computational cost, enabling efficient exploration of kilonova emission.

\section{Acknowledgments}


DB is partially supported by a NASA Future Investigators in NASA Earth and Space Science and Technology (FINESST) award No. 80NSSC23K1440 and the Hearts to Humanity Eternal Research Grant through UC Berkeley. R.M. acknowledges support by the National Science Foundation under award No.  AST-2224255. 
DK is supported in part by the U.S. Department of Energy, Office of Science, Division of Nuclear Physics, under award numbers DE-SC0004658 and DE-SC0024388, and by the Simons Foundation (award number 622817DK).
This research used the Savio computational cluster resource provided by the Berkeley Research Computing program at the University of California, Berkeley (supported by the UC Berkeley Chancellor, Vice Chancellor for Research, and Chief Information Officer).


%



\software{numpy \citep{Numpy}, \texttt{sedona} \citep{Kasen06,Roth15}, 
matplotlib \citep{Matplotlib}, h5py
          }




\appendix

\begin{figure*}
    \centering
    \includegraphics[width=0.90\textwidth,trim={0cm 0.3cm 0cm 0.2cm},clip]{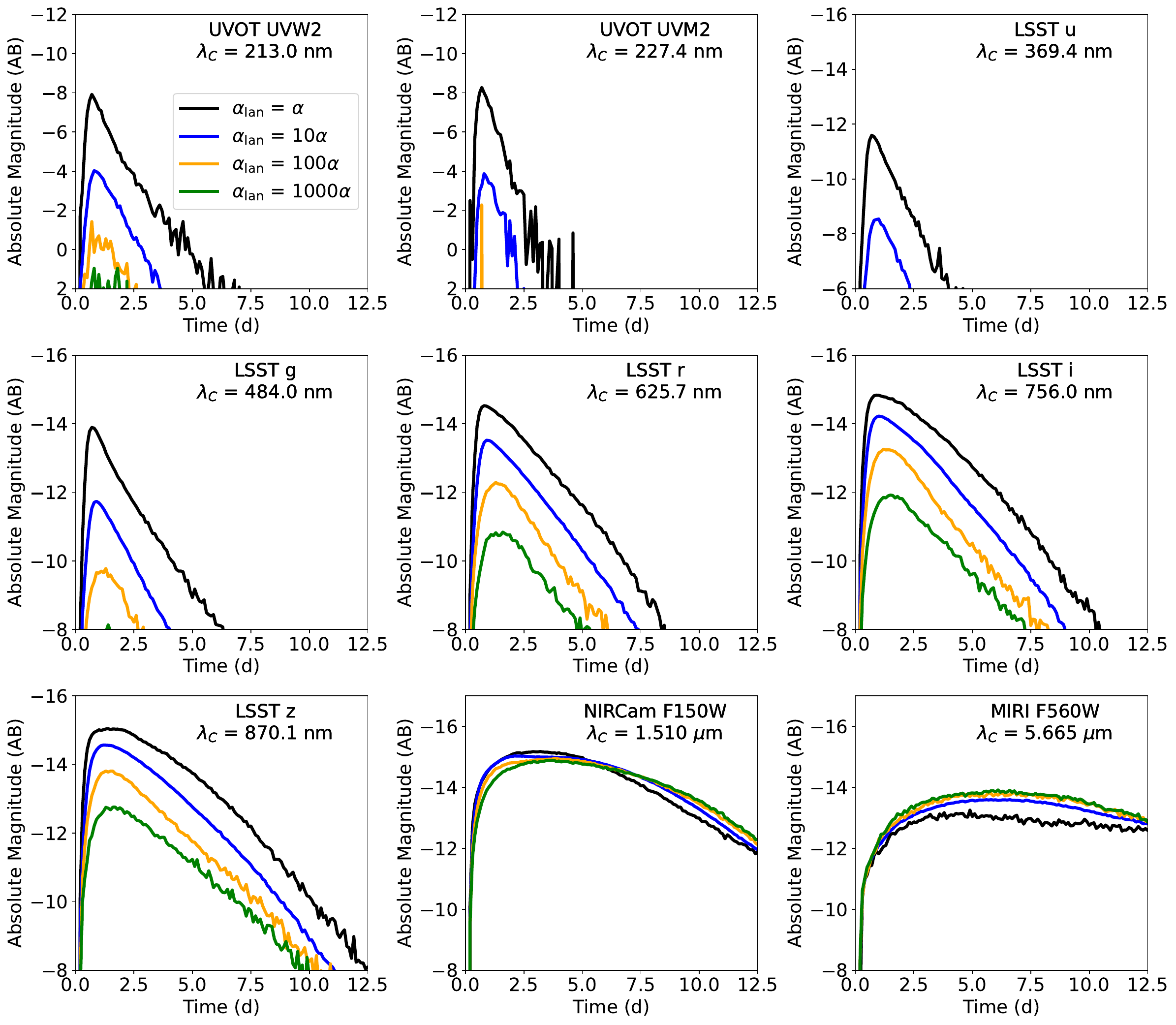}
    \caption{Dependence of the broadband light curves on the lanthanide recombination rate of the ejecta in QNLTE models. The bluest filters are most strongly affected as each light curve is shifted to fainter magnitudes. The IR emission is slightly higher to compensate the decrease in optical emission for higher recombination rates. 
    }
    \label{Fig:AlphaLightCurve}
\end{figure*}

\section{Recombination Rate Uncertainty} \label{App:Recomb}
The recombination rate of lanthanide elements is highly unconstrained due to uncertainty in the strength of dielectronic recombination. Dielectronic recombination may dominate over radiative recombination by more than an order of magnitude in some cases \citep{Hotokezaka21,Singh25} and may also impact light $r$-process elements, though less severely \citep{Banerjee25}. However, there have not been any systematic calculations of the dielectronic recombination rates for lanthanides that can be utilized for radiative transfer simulations. We attempt to bracket the uncertainty in the dielectronic recombination rate by enhancing the recombination rate of lanthanides by a factor of 100 and 1000, and compare the light curves to our fiducial model and an unenhanced recombination rate model (Fig. \ref{Fig:AlphaLightCurve}). As the recombination rate enhances, the kilonova is fainter at blue wavelengths and subtly brighter at IR wavelengths due to less ionized material present. Beyond the overall dimming at UV and optical wavelengths, the shape of the light curves remain remarkably similar over three orders of magnitude in recombination rate with only a modest flattening of the peak visible at the highest recombination rate.



\bibliography{KNe}{}
\bibliographystyle{aasjournal}



\end{document}